\documentclass[reprint, aps,pre,superscriptaddress,showpacs, showkeys]{revtex4-1}

\usepackage[english]{babel}
\usepackage{times}
\usepackage{amssymb}
\usepackage{amsmath}
\usepackage{epsfig}
\usepackage{graphicx}
\usepackage{dcolumn}
\usepackage{color}
\usepackage{bm}

\begin{document}
\bibliographystyle{apsrev4-1}
\pagenumbering{roman}


\title{Quantifying multivariate redundancy with maximum entropy decompositions \\of mutual information}

\author{Daniel Chicharro}
 \affiliation{Department of Neurobiology, Harvard Medical School, Boston, MA, USA}
 \affiliation{Neural Computation Laboratory, Center for Neuroscience and Cognitive Systems@UniTn, Istituto Italiano di Tecnologia, Rovereto (TN), Italy.}%
 \email{daniel\_chicharro@hms.harvard.edu or daniel.chicharro@iit.it}


\begin{abstract}

Williams and Beer (2010) proposed a nonnegative mutual information decomposition, based on the construction of redundancy lattices, which allows separating the information that a set of variables contains about a target variable into nonnegative components interpretable as the unique information of some variables not provided by others as well as redundant and synergistic components. However, the definition of multivariate measures of redundancy that comply with nonnegativity and conform to certain axioms that capture conceptually desirable properties of redundancy has proven to be elusive. We here present a procedure to determine nonnegative multivariate redundancy measures, within the maximum entropy framework. In particular, we generalize existing bivariate maximum entropy measures of redundancy and unique information, defining measures of the redundant information that a group of variables has about a target, and of the unique redundant information that a group of variables has about a target that is not redundant with information from another group. The two key ingredients for this approach are: First, the identification of a type of constraints on entropy maximization that allows isolating components of redundancy and unique redundancy by mirroring them to synergy components. Second, the construction of rooted tree-based decompositions of the mutual information, which conform to the axioms of the redundancy lattice by the local implementation at each tree node of binary unfoldings of the information using hierarchically related maximum entropy constraints. Altogether, the proposed measures quantify the different multivariate redundancy contributions of a nonnegative mutual information decomposition consistent with the redundancy lattice.

\end{abstract}

\maketitle

\section{Introduction}
\label{s1}

Understanding how it is distributed among the components of a multivariate system the information about an external variable, or the reciprocal information of its parts, can help to characterize and to infer the underlying mechanisms and function of the system. This objective has motivated the introduction of different methods to break down the components of the joint entropy of a set of variables \citep{Amari16, Schneidman03b} or to break down the contributions of a set of variables to the mutual information about a target variable \citep{Timme14}. These methods have many applications to study complex systems in the biological domain, such as genes networks \citep[e.\,g.\,][]{Watkinson09, Erwin09, Chatterjee16}, or neural coding \citep[e.\,g.\,][]{Panzeri08, Marre09, Oizumi14, Faes16, Panzeri17}, as well as in the social domain, such as collective behaviour \citep[e.\,g.\,][]{Katz11, Daniels16} and decision agents \citep[e.\,g.\,][]{Polani11}, or to study artificial agents \citep[e.\,g.\,][]{Ay2012}.

In particular, consider a target set of variables $X$ and another set of variables $S$ which information about $X$ we want to study. An important question to determine how this information is distributed refers to how much information is redundant across the variables in $S$, or alternatively can only be obtained synergistically, that is, from the joint observation of the variables \citep{Schneidman03}. The amount of redundant or synergistic information has implications for example to assess how robust the representation of $X$ is \citep{Rauh14b}, how complex it is to decode the information \citep{Latham05}, or how we can reduce the dimensionality of $S$ preserving the information about $X$ \citep{Tishby99}. Several decompositions have been proposed to address this question by breaking down the total information based on the role of correlations between the variables in $S$ not explained by $X$ \citep{Panzeri99b, Chicharro14b}, and more generally separating the influence of dependencies of different orders using maximum entropy models \citep{Amari01, Ince10, Perrone16}. In these models, synergy can be associated with the information that can only be retrieved when considering high-order moments of the joint distribution of the variables. Conversely, redundancy has traditionally been quantified with the measure called interaction information \citep{McGill54} or co-information \citep{Bell03}. However, this measure cannot separate the effect of redundancy and synergy, and which one is predominant is related to the sign of the measure.

As a framework to jointly quantify synergy and redundancy, the seminal work of \cite{Williams10} introduced a new approach to decompose the mutual information that the variables in $S$ (primary sources) have about the target $X$ into a set of nonnegative contributions that differentiate synergy and redundancy. In its simplest bivariate formulation, the mutual information is decomposed into four terms: A redundancy component between variables $1$ and $2$, two terms corresponding to the information that is unique of $1$ and of $2$, respectively, that is, some information that can be obtained from one of the variables alone but not from the other alone, and a synergy term corresponding to the information that is unique for the joint source $12$ with respect to the variables alone. This decomposition not only separates redundancy and synergy but consistently leads to express the measure of co-information as the difference between the redundancy and the synergy terms. This framework more generally allows building this type of decomposition for any multivariate set of variables $S$ \citep{Williams10}. The linchpin ingredients are the definition of a general measure of redundancy that fulfills a set of axioms that capture the abstract notion of redundancy \citep{Williams10b}, and the construction of a redundancy lattice that reflects the partial ordering of different redundancy terms which results from the axioms \citep{Williams10}.

Different elements of this framework have gathered different degrees of consensus. The separation of mutual information into nonnegative components that differentiate redundancy and synergy has been adopted by many others \citep[e.\,g.\,][]{Harder12, Bertschinger12, Griffith13}, but the definition of the measures of redundancy is still a topic of ongoing research. It has been argued that the original redundancy measure of \cite{Williams10} quantifies common amounts of information and not qualitatively common information that the sources share about the target \citep{Harder12}. As a condition to ensure that qualitative redundancy is captured, a new axiom named identity axiom was formulated \citep{Harder12}. New measures that fulfill this axiom have been proposed to underpin the decomposition: From the alternative proposals, some take as the basic component to derive the terms in the decomposition another measure of redundancy \citep{Harder12, Rauh17}, or a measure of synergy \citep{Griffith13}, or of unique information \citep{Bertschinger12, James17}. Moreover, subsequent studies have pointed out new candidate properties to be fulfilled \citep{Bertschinger12b, Griffith13, Harder12, Rauh14, Griffith14, Rauh17b} and there is currently no consensus on which are the properties that should be imposed and on whether there is a preeminent measure of redundancy or several respond to complementary irreducible notions. Only for specific systems such as Gaussian systems with univariate targets, it has been shown that several of the proposed measures are actually equivalent \citep{Barret15, Faes17}. On the other hand, some new proposals depart more substantially from the original framework, either by adopting new principles \citep{Quax17}, by considering the existence of negative components \citep{Ince16, Finn18} associated with misinformation \citep{Wibral14b}, or by implementing decompositions of the joint entropy instead of the mutual information \citep{Rosas16, Ince17}.

The difficulties to decompose mutual information into redundant and synergistic components have been so far substantially aggravated in the multivariate case. While the measures of synergy are more easily generalized to the multivariate case, in particular within the maximum entropy framework \citep{Griffith13, Olbrich15, Perrone16}, the new redundancy measures that have been proposed are only defined for the bivariate case \citep{Harder12, Bertschinger12, Rauh17}, or allow negative components \citep{Ince16, Finn18}. The work in \cite{Chicharro17} introduced two equivalent procedures to construct multivariate redundancy measures, either exploiting the connections between lattices formed by different number of variables, or exploiting the dual connection between redundancy lattices and information loss lattices, for which synergy is more naturally defined. However, these procedures, as in general the framework of \cite{Williams10}, do not guarantee by construction the nonnegativity of the multivariate redundancy measures, which has to be assessed separately for each specific measure.

Furthermore, \cite{Bertschinger12b, Rauh14} showed, with a counterexample involving a trivariate system with deterministic dependencies between the target and the primary sources, that nonnegativity is not ensured for the terms of the redundancy lattice when imposing the identity axiom. The work in \cite{Chicharro17c} generally studied the effect of deterministic target-source dependencies. They indicated how in general negative terms can originate from these dependencies when the measures used to build the decomposition comply with the redundancy axioms of \cite{Williams10} and with an information identity criterion that subsumes the identity axiom and generalizes to the multivariate case. This criterion assumes that different pieces of information in the target can be specifically associated with different source variables. It remains an open question whether the lack of a nonnegative decomposition is a signature of certain target-source dependencies or if meaningful nonnegative decompositions could be obtained replacing the identity criterion.

In this work we propose a procedure to determine multivariate redundancy measures within the maximum entropy framework, thus extending the bivariate approach of \cite{Bertschinger12}. This focus on the maximum entropy approach is motivated by its preeminent role for the bivariate case, where it provides bounds for the actual redundancy, unique information, and synergy terms under reasonable assumptions shared by other measures \citep{Williams10, Harder12}. In particular, \cite{Bertschinger12} showed that, if it is assumed that a bivariate nonnegative decomposition exists and that redundancy can be determined from the bivariate distributions of the target with each source, then the maximum entropy measures provide a lower bound for the actual synergy and redundancy terms, and an upper bound for the actual unique information. Furthermore, under the assumptions above, if these bivariate distributions are compatible with potentially having no synergy (see details in Section \ref{s2_2}), then the maximum entropy decomposition retrieves not only bounds but the exact actual terms. We here show that this preeminent role of the maximum entropy decomposition also holds for the multivariate case under analogous assumptions.

To construct maximum entropy multivariate mutual information decompositions we define measures of the redundancy that a set of variables have about $X$, and also of the unique redundant information that a set of variables have about $X$ that is not redundant with another set of variables. The formulation within the maximum entropy framework allows obtaining close-form general definitions of the multivariate redundancy measures, in contrast to \cite{Chicharro17}, and ensures the nonnegativity of the measures by construction, imposing hierarchical constraints to entropy maximization. The key ingredient to define these measures is the identification of a type of entropy maximization constraint using co-information measures, which allows separating the components of the multivariate redundancy. When all terms in the mutual information decomposition are nonnegative, the constraints used to define the different measures are consistent across measures, and the multivariate redundancy measures implement a decomposition that breaks down the redundancy contributions of the mutual information in agreement with the redundancy axioms of \cite{Williams10}. In particular, the measures are related to each other forming a rooted tree decomposition in which different redundancy terms are locally binary unfolded using hierarchically related maximum entropy constraints. Oppositely, for those systems with target and source dependencies that do not allow a nonnegative mutual information decomposition compatible with the maximum entropy approach \citep{Chicharro17c}, we show that the hierarchy of the constraints is lost. In that case, while the measures can still be calculated and are nonnegatively defined, their consistency as a mutual information decomposition does no longer hold, so that their interpretability as quantifying the different redundancy contributions is impaired.

In Section \ref{s2} we revise the principles of the mutual information decomposition \citep{Williams10} and the bivariate maximum entropy measures \cite{Bertschinger12}. In Section \ref{s3} we revisit these measures to make more apparent how they can be generalized to the multivariate case. In particular, we identify how entropy maximization constraints on co-information allow isolating the redundancy components. In Section \ref{s4} we develop the general multivariate measures. We indicate that the preeminent role of the maximum entropy decomposition holds also for the multivariate case (Section \ref{s4_0}). We start with the trivariate case (Section \ref{s4_1}) and then present the general development of the multivariate redundancy decompositions (Section \ref{s4_2}). We provide general formulas of the effect of constraints on co-information and conditional co-information and use them to derive multivariate measures of redundancy and unique redundancy. We then show that these measures can implement a nonnegative rooted tree decomposition of the mutual information via local binary unfoldings of the information at each tree node using hierarchically related maximum entropy constraints. In Section \ref{s5} we show how the presence of negative terms impairs the consistency of the decomposition by breaking the hierarchy of these constraints.

\section{A review of lattice-based decompositions of mutual information}
\label{s2}

The seminal work of \cite{Williams10, Williams10b} introduced a new approach to decompose mutual information into a set of nonnegative contributions. Let us consider first the bivariate case, with a target variable $X$ and two source variables $1$ and $2$. The work in \cite{Williams10} showed that the mutual information of each variable can be expressed as
\begin{equation}
I(X;1) = I(X;1.2)+I(X;1 \backslash 2),
\label{e1}
\end{equation}
and similarly for $I(X;2)$. The term $I(X;1.2)$ refers to a redundancy component between variables $1$ and $2$, which can be obtained either by knowing $1$ or $2$ separately. The terms $I(X;1 \backslash 2)$ and $I(X;2 \backslash 1)$ quantify a component of the information that is unique of $1$ and of $2$, respectively, that is, the information that can be obtained from one of the variables alone but that cannot be obtained from the other alone. Furthermore, the joint information of $12$ can be expressed as
\small
\begin{equation}
I(X;12) =I(X;1.2)+I(X;1 \backslash 2)+I(X;2 \backslash 1)+I(X;12 \backslash 1,2),
\label{e2}
\end{equation}
\normalsize
where $I(X;12 \backslash 1,2)$ refers to the synergistic information of the two variables, which is unique for the joint source $12$ with respect to both variables alone. Therefore, given the standard information-theoretic chain rule equalities \citep{Cover06}
\begin{subequations}
\begin{align}
I(X;12) &= I(X;1)+I(X;2|1)\\
&= I(X;2)+I(X;1|2),
\end{align}
\label{e3}
\end{subequations}
the conditional mutual information is decomposed as
\begin{equation}
I(X;2|1) = I(X;2 \backslash 1)+I(X;12 \backslash 1,2),
\label{e4}
\end{equation}
and analogously for $I(X;1|2)$. Conditioning on one variable removes the redundant component of the information but adds the synergistic component, resulting in the conditional information being the sum of the unique and synergistic terms.

Note that in this decomposition a redundancy and a synergy component can exist simultaneously. In fact, \cite{Williams10} showed that the measure of co-information \citep{Bell03} that previously had been used to quantify synergy and redundancy, defined as
\small
\begin{equation}
C(X;1;2) = I(i;j)-I(i;j|k) = I(i;j)+I(i;k)-I(i;j,k)
\label{e5}
\end{equation}
\normalsize
for any assignment of $\{X,1,2\}$ to $\{i,j,k\}$, equals the difference between the redundancy and synergy terms of Eq.\,\ref{e2}:
\begin{equation}
C(X;1;2) = I(X;1.2)-I(X;12 \backslash 1,2).
\label{e5_2}
\end{equation}

\subsection{Redundancy measures and lattices}
\label{s2_1}

More generally, \cite{Williams10, Williams10b} defined decompositions of the mutual information about a target $X$ for any multivariate set of variables $S$. The key ingredient was the definition of a general measure of redundancy and the construction of a redundancy lattice. We now summarize the elements of their approach relevant for the development of the rooted tree decompositions.

To decompose the information $I(X;S)$, \cite{Williams10} defined a \emph{source} $A$ as a subset of the variables in $S$, and a \emph{collection} $\alpha$ as a set of sources. They then introduced a measure of redundancy to quantify for each collection the redundancy between the sources composing the collection, and constructed a redundancy lattice which reflects the relation between the redundancies of all different collections. Here we will generically refer to the redundancy of a collection $\alpha$ by $I(X;\alpha)$. Furthermore, following \cite{Chicharro17}, we use a more concise notation than in \cite{Williams10}: For example, instead of writing $\{1\}\{23\}$ for the collection composed by the source containing variable $1$ and the source containing $2$ and $3$, we write $1.23$, that is, we save the curly brackets that indicate sets of variables and we use instead a dot to separate the sources. \cite{Williams10b} argued that a measure of redundancy should comply with the following axioms:

\begin{itemize}
  \item \textbf{Symmetry}: $I(X;\alpha)$ is invariant to the order of the sources in the collection.
  \item \textbf{Self-redundancy}: The redundancy of a collection formed by a single source is equal to the mutual information of that source.
  \item \textbf{Monotonicity}: Adding sources to a collection can only decrease the redundancy of the resulting collection, and redundancy is kept constant when adding a superset of any of the existing sources.
\end{itemize}

\begin{figure}
  \begin{center}
    \scalebox{0.34}{\includegraphics*{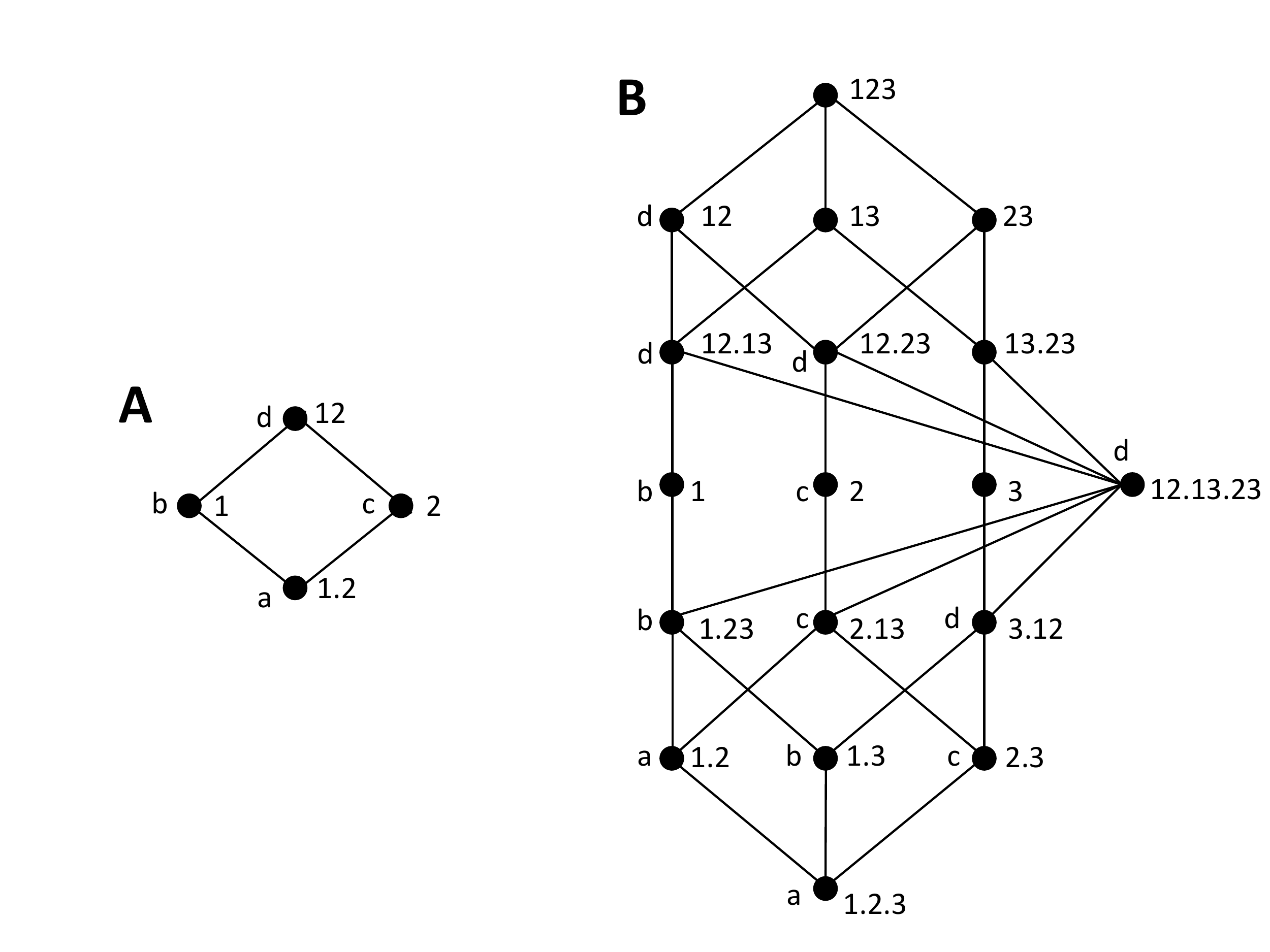}}
  \end{center}
  \caption{Bivariate and trivariate redundancy lattices of \cite{Williams10}. The lattices reflect the partial ordering defined by Eq.\,\ref{e7}. The labels indicate the mapping from the trivariate to the bivariate lattice. In particular, nodes with the same label in the trivariate lattice are accumulated in the corresponding node in the bivariate lattice.}
  \label{fig0}
\end{figure}

The monotonicity property allows introducing a partial ordering between the collections, which is reflected in the redundancy lattice. Self-redundancy connects the lattice to a decomposition of the mutual information $I(X;S)$ because at the top of the lattice there is the collection formed by a single source including all the variables in $S$. Furthermore, the number of collections to be included in the lattice is restricted by the fact that adding a superset of any source does not change redundancy. For example, the redundancy between sources $12$ and $1$ is all the information $I(X;1)$. Accordingly, the set of collections included in the lattice is defined as
\begin{equation}
\mathcal{A}(S)= \{ \alpha \in \mathcal{P}(S) - \{ \emptyset \}: \forall\ A_i, A_j \in \alpha, A_i \nsubseteq A_j\},
\label{e6}
\end{equation}
where $\mathcal{P}(S) - \{ \emptyset \}$ is the set of all nonempty subsets of the set of nonempty sources that can be formed from $S$. This domain reflects the symmetry axiom in that it does not distinguish the order of the sources. For this set of collections, \cite{Williams10} defined a partial ordering relation to construct the lattice:
\begin{equation}
\forall\ \alpha, \beta \in \mathcal{A}(S), (\alpha \preceq \beta \Leftrightarrow \forall B \in \beta, \exists A \in \alpha, A \subseteq B),
\label{e7}
\end{equation}
that is, for two collections $\alpha$ and $\beta$, $\alpha \preceq \beta$ if for each source in $\beta$ there is a source in $\alpha$ that is a subset of that source. This partial ordering relation is reflexive, transitive, and antisymmetric. In fact, the consistency of the redundancy measures with the partial ordering of the collections, that is, that $I(X;\alpha) \leq I(X;\beta)$ if $\alpha \preceq \beta$ represents a stronger condition than the monotonicity axiom. This is because the monotonicity axiom only considers the cases in which $\alpha$ is obtained from $\beta$ adding more sources (e.g., $\alpha= 1.2.3$ and $\beta= 1.2$), while the partial ordering comprises also the removal of variables from sources (e.g., $\alpha= 1.2$ and $\beta= 1.23$, or $\alpha= 1$ and $\beta= 12.13$). The redundancy lattices for the case of $S$ being bivariate and trivariate are shown in Figure \ref{fig0} indicating their correspondence \citep{Chicharro17}. The three axioms above and the partial ordering relations reflected in the lattices will also underpin the multivariate redundancy measures proposed in this work.

The mutual information decomposition was constructed in \cite{Williams10} by implicitly defining partial information measures $\Delta_{\mathcal{C}}(X;\alpha)$ associated with each node $\alpha$ of the redundancy lattice $\mathcal{C}$, such that the redundancy measures are obtained as
\begin{equation}
I(X;\alpha) = \sum_{\beta \in \downarrow \alpha} \Delta_{\mathcal{C}}(X;\beta),
\label{e8}
\end{equation}
where $\downarrow\alpha$ refers to the set of collections lower than or equal to $\alpha$ in the partial ordering, and hence reachable descending from $\alpha$ in the lattice $\mathcal{C}$. The decomposition of the total mutual information results from applying Eq.\,\ref{e8} to the collection $\alpha = S$. The decompositions of Eqs.\,\ref{e1} and \ref{e2} are particular cases of Eq.\,\ref{e8}. The partial information measures are obtained by inverting Eq.\,\ref{e8}, applying the M{\"o}bius inversion \citep{Williams10}. In the bivariate case the partial information measures can be identified as redundant, unique, and synergistic terms, respectively. Oppositely, in the multivariate case, the partial information measures quantify in general contributions which result from a mixture of these notions and represent in general the part of the redundancy that is unique for one collection with respect to others. As we will show, this idea of separating unique and common parts of the redundancy is at the core of the tree decompositions.

\subsection{Bivariate maximum entropy decompositions of mutual information consistent with the redundancy lattice}
\label{s2_2}

The work in \cite{Harder12} argued that the original measure of redundancy \cite{Williams10} is not suited because it only quantifies common amounts of information, and introduced a new axiom as a necessary condition to qualitatively quantify redundancy:

\begin{itemize}
  \item \textbf{Identity axiom:} For two sources $A_1$ and $A_2$,  $I(A_1 \cup A_2; A_1.A_2)$ is equal to $I(A_1;A_2)$.
\end{itemize}

This new axiom has motivated the proposal of several alternative measures of redundancy \citep[e.\,g.\,][]{Harder12, Bertschinger12, Griffith13, Ince16, James17}. From these measures, we focus on the maximum entropy-based redundancy measure of \cite{Bertschinger12} and the associated mutual information decomposition, and we will develop the rooted tree decomposition as its multivariate extension. This is motivated by the preeminent role of the bivariate maximum entropy decomposition in relation to the underlying actual decomposition \cite{Bertschinger12}, which we will show that also holds for the multivariate case. The bivariate decomposition of \cite{Bertschinger12} complies with the identity axiom, and also with its generalization introduced in \cite{Chicharro17c}.

The maximum entropy measures are defined by comparing the mutual information $I(X;12)$ that variables $1$ and $2$ have about $X$ with the minimum mutual information for a set of joint distributions $q(X,1,2)$ defined as:
\begin{equation}
\Delta_{1.2}(p) = \left \{q : q(x,1)= p(x,1), q(x,2)= p(x,2) \right \},
\label{e9}
\end{equation}
that is, the set of distributions that preserve the bivariate marginals involving $X$, which also includes the original distribution $p$. It is then assumed that unique information and redundancy are invariant within this set, while synergy depends on the specific form of the trivariate joint distributions. Synergy is determined as the difference between the information for the original $p(X,1,2)$ and the minimum within $\Delta_{1.2}(p)$:
\begin{equation}
\begin{split}
I(X;12 \backslash 1,2) &= I(X;12)-\min \limits_{1.2} I(X;12) \\&= \max \limits_{1.2} H(X|12) -H(X|12),
\label{e10}
\end{split}
\end{equation}
where we abbreviate $q \in \Delta_{1.2}(p)$ to ${1.2}$ for the minimization constraint. The formulation in terms of maximum entropy is possible because $H(X)$ is preserved in $\Delta_{1.2}(p)$. According to this maximum entropy interpretation, the synergy is nonnegatively defined. Using the relations of Eqs.\,\ref{e1}-\ref{e4}, the other components of the decomposition are derived to be:
\begin{subequations}
\begin{align}
I(X;2 \backslash 1) &= \min \limits_{1.2} I(X;2|1)\\
I(X;1 \backslash 2) &= \min \limits_{1.2} I(X;1|2) \\
I(X;1.2) &= I(X;1)+I(X;2)- \min \limits_{1.2} I(X;12) \\
&= I(X;1)-\min \limits_{1.2} I(X;1|2) \notag \\
&= \max \limits_{1.2} C(X;1;2). \notag
\end{align}
\label{e11}
\end{subequations}
\cite{Bertschinger12} proved that these measures compose a nonnegative decomposition of the mutual information. The nonnegativity of the synergistic and unique information measures is apparent from their expressions above, while for redundancy it can be checked by showing that at least there is a distribution within $\Delta_{1.2}(p)$ for which the co-information is nonnegative.

Furthermore, these maximum entropy measures impose certain bounds to the underlying actual terms of redundancy, synergy and unique information that they estimate \citep{Bertschinger12}. In particular, under the assumptions that the actual decomposition exists, that the actual synergy is nonnegative, and that the actual redundancy can be determined from the bivariate distributions of the target with each source, then $I_0(X;1.2) \geq I(X;1.2)$, $I_0(X;12 \backslash 1,2) \geq I(X;12 \backslash 1,2)$, and $I_0(X;i \backslash j) \leq I(X;i \backslash j)$, where $I_0$ indicates the actual terms. These bounds become equalities if there is one distribution within $\Delta_{1.2}(p)$ for which $I_0(X;12 \backslash 1,2)=0$ \cite{Bertschinger12}. In that case, the maximum entropy measures retrieve exactly the actual decomposition and can be interpreted as following: Eq.\,\ref{e4} indicates that the conditional mutual information decomposes into a unique and a synergistic component. Since the distributions in $\Delta_{1.2}(p)$ are not constrained to maintain the joint distribution $p(X,1,2)$, the minimization gets rid of the synergy, selecting distributions with $I_0(X;12 \backslash 1,2)=0$. This allows getting the unique information as equal to the conditional information. Since given Eq.\,\ref{e1} the mutual information of each variable is the sum of the redundancy and unique component, the redundancy is then retrieved by the mutual information minus the conditional. We will show that this especial relation of the actual decomposition with its maximum entropy estimation also holds for the multivariate case.

\begin{figure}
  \begin{center}
    \scalebox{0.37}{\includegraphics*{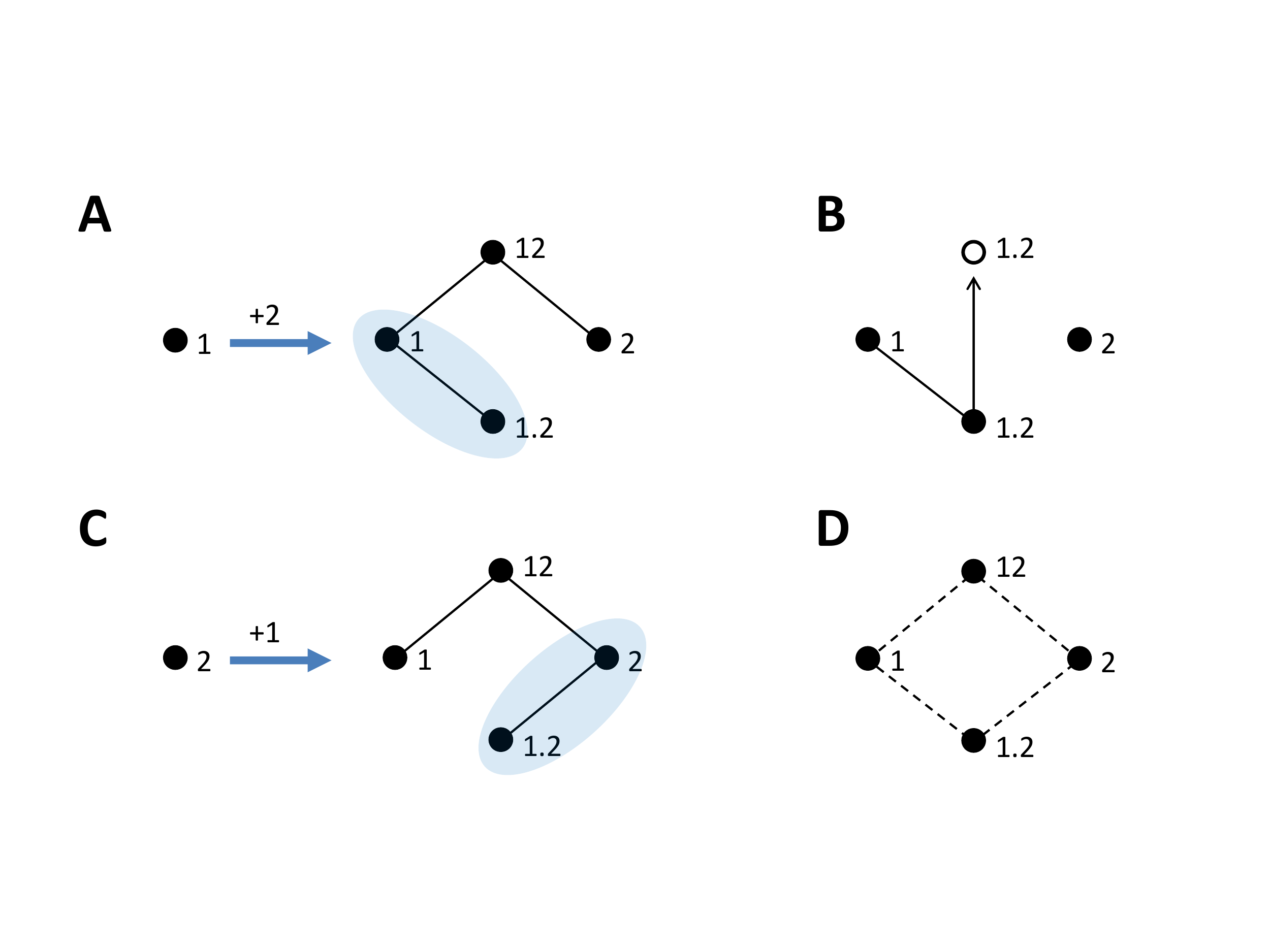}}
  \end{center}
  \caption{Rooted trees associated with the decomposition of the mutual information that variables $1$ and $2$ have about target $X$. \textbf{A)} Tree formed by adding variable $2$ after variable $1$. The shaded elliptical area indicates the unfolding of information already present previous to adding the new variable into an information component redundant to the new variable and a component unique with respect to it. \textbf{B)} Graphical representation of the effect of a constraint imposing a null co-information when preserving only the marginal distributions of the variables with the target. The arrow indicates that redundancy is mirrored to the otherwise minimized synergistic component (empty circle). \textbf{C)} Same as A) for the permute order of addition of the variables. \textbf{D)} Adjunct lattice to the rooted trees. This lattice has the same structure as the bivariate redundancy lattice of \cite{Williams10} and each edge reflects the existence of that edge in at least one of the two trees.}
  \label{fig1}
\end{figure}

\section{Bivariate rooted tree-based maximum entropy decompositions of mutual information }
\label{s3}

As a first step towards defining the multivariate redundancy decompositions, we first revisit the bivariate maximum entropy decompositions of \cite{Bertschinger12} interpreting them as associated with rooted trees. For that purpose, consider that we start by having access only to one of the two variables, e.\,g.\,variable $1$. Accordingly, we only have the information $I(X;1)$, and there are no other variables to establish which part is redundant or unique. After adding $2$, $I(X;1)$ unfolds into two components, as indicated by the shaded ellipse of Figure \ref{fig1}A: The information of $1$ can be split between the component redundant to $2$ and the component unique with respect to $2$. Furthermore, there is new information, which can itself be separated into two components: The unique information of $12$ with respect to $1$ that is redundant with the unique information of $2$ with respect to $1$, and the unique information of $12$ with respect to $1$ that is also unique with respect to the unique information of $2$ with respect to $1$.

Therefore, the four terms of the decomposition result from two separate unfoldings: Of the information $I(X;1)$, already present previous to the addition of $2$, and of the new information $I(X;2|1)$, respectively. Accordingly, to guarantee that the terms of the decomposition are nonnegative, we can do it locally, that is, we only have to check that the binary unfoldings are into nonnegative components. To see how this works we start reexpressing the measures of Eqs.\,\ref{e10}
and \ref{e11}:
\begin{subequations}
\begin{align}
I(X;12 \backslash 1,2) &= I(X;12)-\min \limits_{1.2} I(X;12)\\
I(X;2 \backslash 1) &= \min \limits_{1.2} I(X;12)-\min \limits_{1.2} I(X;1)\\
I(X;1 \backslash 2) &=  \min \limits_{1.2} I(X;12)-\min \limits_{1.2} I(X;2)\\
I(X;1.2) &= \min \limits_{1.2, C(X;1;2)=0} I(X;12)-\min \limits_{1.2} I(X;12).
\end{align}
\label{e12}
\end{subequations}
The synergy is expressed as before, as a comparison of the information for the original distribution and the minimum information for $\Delta_{1.2}(p)$. The two unique informations are expressed as a difference between two informations within $\Delta_{1.2}(p)$. The equivalence with the previous expressions holds because the marginal informations $I(X;1)$ and $I(X;2)$ are preserved within $\Delta_{1.2}(p)$. The redundancy is now formulated as a comparison of informations for two hierarchical minimizations, and hence its nonnegativity is directly apparent, in contrast to in Eq.\,\ref{e11}c. In particular, we introduce the set
\begin{equation}
\begin{split}
\Delta_{1.2, C(X;1;2)}(p) = &\left\{ q : q(x,1)= p(x,1), q(x,2)= p(x,2) \right. ,\\ & \left. C(X;1;2) =0 \right\}.
\label{e13}
\end{split}
\end{equation}
The equivalence to Eq.\,\ref{e11}c can be checked using the definition of co-information (Eq.\,\ref{e5}) and the equality resulting from $C(X;1;2) =0$. Nonnegativity is guaranteed by the minimizations within sets such that $\Delta_{1.2, C(X;1;2)}(p) \subseteq \Delta_{1.2}(p)$. Figure \ref{fig1}B provides some intuition about how redundancy is quantified: as mentioned above, $\min \limits_{1.2} I(X;12)$ tends to eliminate the synergy component of the original distribution. Conversely, redundancy is constant within $\Delta_{1.2}(p)$. Therefore, imposing that the co-information vanishes, since it is equal to redundancy minus synergy (Eq.\,\ref{e5_2}), enforces to select those distributions for which the synergy component is not eliminated, but mirrors the redundancy component.

Given the reexpressed measures, it is straightforward to check how the unfoldings are implemented:
\small
\begin{equation}
\begin{split}
I(X;1) &= \min \limits_{1.2, C(X;1;2)=0} I(X;12)-\min \limits_{1.2} I(X;2)\\
&= \left [\min \limits_{1.2, C(X;1;2)=0} I(X;12)-\min \limits_{1.2} I(X;12)\right ]\\&+  \left [\min \limits_{1.2} I(X;12)-\min \limits_{1.2} I(X;2) \right ]
\end{split}
\label{e14}
\end{equation}
\normalsize
and
\small
\begin{equation}
\begin{split}
I(X;2|1) &= I(X;12)-\min \limits_{1.2} I(X;1)\\
&= \left [I(X;12)-\min \limits_{1.2} I(X;12)\right ] \\&+ \left [\min \limits_{1.2} I(X;12)-\min \limits_{1.2} I(X;1) \right ].
\end{split}
\label{e15}
\end{equation}
\normalsize
Furthermore, it can be immediately checked that the measures are also consistent with the reverse ordering of addition of the variables (Figure \ref{fig1}C), in which case the pairs resulting from each unfolding are interchanged. This allows associating the trees with a common adjunct lattice equivalent to the one defined by \cite{Williams10} (Figure \ref{fig1}D). An edge in this lattice indicates an edge that is present in at least one of the trees.

We can compare this maximum entropy decomposition with other approaches based on maximum entropy \citep{Amari01, Ince10, Olbrich15, Perrone16}. The construction of the synergistic term in Eq.\,\ref{e12}a is equivalent to the usual maximum entropy decompositions in which marginal distributions up to a certain order are preserved. However, this same strategy cannot identify redundancy terms because they are embedded in the information kept by maintaining only the bivariate marginals involving $X$ and each variable in $S$. The new type of constraint on the co-information allows gaining access to the redundancy terms by mirroring them into the synergistic component, which otherwise would be minimized given that there is no constraint to preserve high-order marginals of the distribution.

\section{Multivariate rooted tree-based maximum entropy decompositions of mutual information }
\label{s4}

We now generalize to the multivariate case the procedure to construct rooted tree maximum entropy decompositions. We first indicate that the preeminent role of bivariate maximum entropy decompositions \citep{Bertschinger12} also holds for the multivariate case (Section \ref{s4_0}). Then, to gain some further intuition, we examine in detail the trivariate case (Section \ref{s4_1}). In Section \ref{s4_2} we introduce the general multivariate redundancy measures and we describe the multivariate decompositions.

\subsection{The relation between the decomposition terms and their maximum entropy estimators}
\label{s4_0}

As described in Section \ref{s2_2}, the bivariate maximum entropy decomposition retrieves the actual one if this decomposition fulfills certain properties. This relation can be extended to the multivariate case. We here enunciate this result and we provide the proof in Appendix \ref{ap}. Consider the following assumption for the actual decomposition:

\vspace*{3mm}

\noindent \textbf{Assumption a.1}: Consider a distribution $p(X,S)$. The distributions in the family $\Delta_{S_1.S_2.....S_n}(p)$ that preserves the marginal distributions $p(X,S_i)$, for each $S_i \in S$ are such that they allow constructing decompositions of the mutual information consistent with the partial ordering of redundancies of Eq.\,\ref{e7}, with the relations between mutual information and redundancy measures of Eq.\,\ref{e8}, and in which all terms are nonnegative. Furthermore, for any $q \in \Delta_{S_1.S_2.....S_n}(p)$, all terms which do not involve synergy in the decompositions of $I_q(X,S')$ with $S'\subseteq S$, are invariant within this family.
\vspace*{3mm}

The assumption on the invariance of nonsynergistic terms within the family is analogous to Assumption $(*)$ of \cite{Bertschinger12}. For example, for the family $\Delta_{1.2.3}(p)$ defined analogously to Eq.\,\ref{e9}, the terms $\Delta(X;i.j)$ and $\Delta(X,i.j.k)$ of the trivariate lattice (Figure \ref{fig0}B) are invariant according to this assumption. Oppositely, the terms $\Delta(X;i.jk)$ are not invariant because they involve the distribution $p(X,j,k)$ not preserved within the family. However, because the assumption also regards any bivariate decomposition related to the trivariate one (Figure \ref{fig0}A), the terms $\Delta(X;i)+\Delta(X;i.jk)$ of the trivariate lattice are also invariant due to the invariance of the terms $\Delta(X;i)$ of the bivariate lattices.

Maximum entropy redundancy decompositions constructed according to Eqs.\,\ref{e7} and \ref{e8} are related to the actual decompositions by the following lemma:
\vspace*{3mm}

\noindent \textbf{Lemma l.1}: \emph{Consider a target $X$ and a set $S$ of $n$ primary sources with a distribution $p(X,S)$. Consider the family of distributions $\Delta_{S_1.S_2.....S_n}(p)$ preserving the marginal distributions of $X$ with each primary source. Assume that an actual decomposition of the mutual information exists for each distribution within the family, conforming to assumption $a.1$. Then:}

\emph{If there is a $q \in \Delta_{S_1.S_2.....S_n}(p)$ for which all the actual synergistic terms between the primary sources vanish, then the redundancy (and unique redundancy) measures of the maximum entropy decomposition retrieve the corresponding redundancy terms of the actual decomposition. Conversely, if the maximum entropy redundancy measures retrieve the corresponding actual redundancy terms, there is a $q \in \Delta_{S_1.S_2.....S_n}(p)$ for which all the actual synergistic terms vanish.}

\vspace*{1.5mm}

\noindent \textbf{Proof.} See Appendix \ref{ap}.
\vspace*{2mm}

This Lemma extends Lemma $3$ of \cite{Bertschinger12}. However, it focuses on the relation between the redundancy and unique redundancy terms of the maximum entropy and actual decompositions because, as we will see, the rooted tree decompositions do not decompose the synergistic contributions. Given the sufficient condition stated in lemma $l.1$ to retrieve the actual terms with the maximum entropy measures, we also adopt the following assumption:
\vspace*{3mm}

\noindent \textbf{Assumption a.2}: The constrained minimization of the mutual information cancels all terms of its actual decomposition whose cancellation is compatible with the minimization constraints.
\vspace*{3mm}

This second assumption considers that within the family of distributions in which we minimize there is a distribution in which all terms of the actual decomposition not affected by the constraints are zero. For example, we assume that, for a minimization within the family of distributions $\Delta_{1.2.3}(p)$, all synergistic terms that cannot be reached descending from the nodes corresponding to primary sources (Figure \ref{fig0}B) are canceled. The combination of assumptions $a.1$ and $a.2$ allows analyzing the effect of minimization constraints only by examining the redundancy lattice, and hence to derive general expressions for the effect of the constraints independently of the specific properties of a certain distribution $p(X,S)$. Accordingly, in the rest of this section we derive our results under $a.1$ and $a.2$. Only in Section \ref{s5} we will examine how the maximum entropy redundancy measures are affected by the break of these assumptions.

\subsection{The trivariate case}
\label{s4_1}

Revisiting the maximum entropy decomposition of \cite{Bertschinger12}, we have introduced a new type of minimization constraint, namely imposing the cancellation of the co-information, which enforces the emergence of redundancy to synergy terms, which are otherwise canceled. The form of these co-information constraints suggests how this approach can be generalized to the multivariate case. However, in the multivariate case different redundancy contributions have to be isolated (Figure \ref{fig0}B) and can only be retrieved by combining different constraints on co-information measures. We now examine how to do so for the trivariate case.

\begin{figure}
  \begin{center}
    \scalebox{0.32}{\includegraphics*{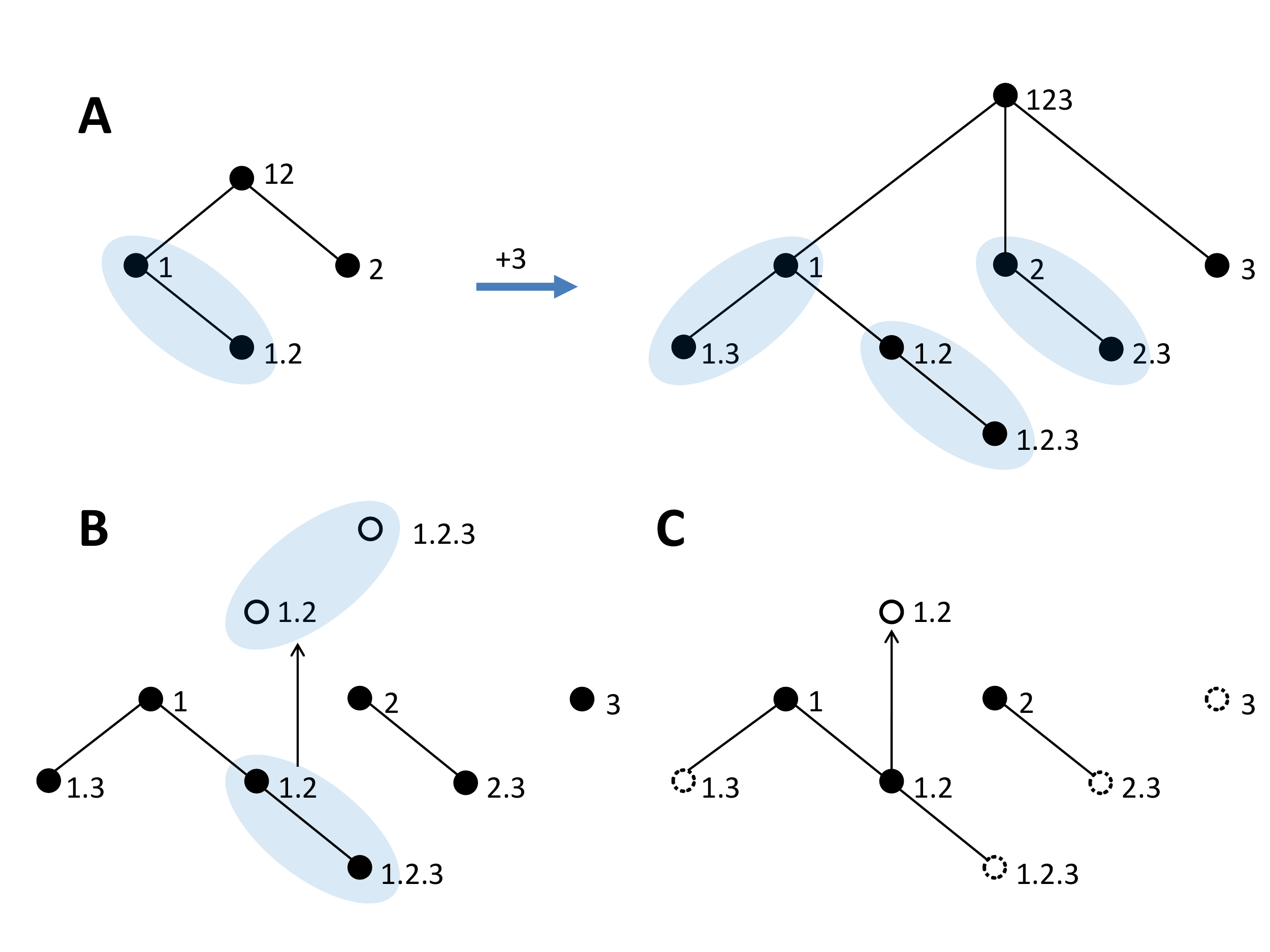}}
  \end{center}
  \caption{Rooted trees associated with the decomposition of the mutual information that variables $1$, $2$, and $3$ have about target $X$. \textbf{A)} Updating of the tree that resulted from adding $2$ after $1$ when $3$ is added subsequently. Like in Figure \ref{fig1}, the elliptical shaded areas indicate the unfolding of a previous node. \textbf{B)} Representation of the effect of the constraint on the unconditional co-information $C(X;1;2)$. The arrow indicates the emergence of redundancy between $1$ and $2$ to the synergistic component accumulated in $123$, which is otherwise minimized (empty circles). \textbf{C)} Representation of the effect of the constraint on the conditional co-information $C(X;1;2|3)$. The arrow indicates the emergence to the synergistic component of the redundancy between $1$ and $2$ unique with respect to $3$. Dashed circles indicate the effect of conditioning on $3$, removing the redundancy components contained in the information that $3$ has about the target.}
  \label{fig2}
\end{figure}

Figure \ref{fig2}A shows how the tree formed by the sequence of additions $1 \rightarrow 2 \rightarrow 3$
is updated from the one of $1 \rightarrow 2$. Again, each redundancy component is unfolded into two new terms. Conversely, the synergistic components are accumulated in the new term $I(X;123 \backslash 1,2,3)$. We will later discuss the implications of this asymmetric updating of redundancy and synergy. We now focus on identifying the terms resulting from the unfolding and checking their nonnegativity. First, we have $I(X;1.2) = I(X;1.2.3)+ I(X; 1.2 \backslash 3)$, that is, the redundancy between $1$ and $2$ unfolds into the redundancy of the three and the redundancy unique of $1$ and $2$ with respect to $3$. This can be seen as a direct consequence of the monotonicity axiom. Second, we have $I(X;1 \backslash  2)= I(X;1 \backslash  2,3)+I(X;1.3 \backslash 2)$, that is, the information of $1$ unique with respect to $2$ is unfolded into the self-redundancy of $1$ unique with respect to both $2$ and $3$ and the redundancy of $1$ and $3$ unique with respect to $2$. The same happens with $I(X;2 \backslash 1)= I(X;2 \backslash  1,3)+I(X;2.3 \backslash  1)$. This can be seen as a consequence of a monotonicity property for unique redundancies analogous to the one of the redundancies.

To see how to introduce constraints that enforce the emergence of different parts of the redundancy by mirroring them to the synergy, we first consider the same constraint used for the bivariate case, that is, $C(X;1;2)=0$. In the bivariate case this constraint imposes $\Delta(X;1.2)-\Delta(X;12) =0$ for the terms of the bivariate lattice (Figure \ref{fig0}A). We now want to find a constraint that leads again to the emergence of $I(X;1.2)$, but that explicitly considers that variable $3$ is involved. As we will see, this is important for the unfolding because we need that the constraints used when including a different number of variables in the tree are hierarchically related. Therefore, we want to take into account that $I(X;1.2)= \Delta(X;1.2) + \Delta(X;1.2.3)$ in the trivariate lattice (Figure \ref{fig0}B). Given that we are only interested in the redundancy between $1$ and $2$, one may think that the way to quantify this redundancy also explicitly considering $3$ could be to use a definition analogous to Eq.\,\ref{e12}d, minimizing on $I(X;123)$ and preserving also the marginal of $3$ with $X$. That is, we could minimize $I(X;123)$ within $\Delta_{1.2.3, C(X;1;2)}(p)$ and $\Delta_{1.2.3}(p)$, respectively. Since only the difference of the minimized informations matters, it would not matter which is their common ground imposed by the constraint common to both minimizations, i.\,e., the preservation of the marginals with the target. However, this is not the case, because the effect of $C(X;1;2)=0$ changes depending on whether only the marginal distributions of $1$ and $2$ with $X$ are preserved, or also the one of $3$ is. To appreciate this, we can write the constraint explicitly, according to Figure \ref{fig0}B:
\begin{equation}
\begin{split}
&C(X;1;2)=0 = I(X;1)-I(X;1|2) \\&=\left [ \Delta(X;1.2) +\Delta(X;1.2.3) \right] \\&- \left [ \Delta(X;12) + \Delta(X;12.13) + \Delta(X;12.23) \right.\\ &\ \ \ \ \ \left. + \Delta(X;12.13.23) + \Delta(X;3.12) \right].
\end{split}
\label{e15_2}
\end{equation}
We now ask how can this constraint be fulfilled while minimizing the mutual information $I(X;123)$ within $\Delta_{1.2.3, C(X;1;2)}(p)$. Following $a.1$, all redundancy and unique information terms are invariant within $\Delta_{1.2.3}(p)$, and hence also in $\Delta_{1.2.3, C(X;1;2)}(p)$, which is subsumed by it. This means that the terms $\Delta(X;1.2)$ and $\Delta(X;1.2.3)$ are fixed and equal to the ones of the original distribution. Furthermore, also $\Delta(X;3)+\Delta(X;3.12)$ is constant, because it corresponds to the unique information of variable $3$ with respect to variables $1$ and $2$. However, while their sum is fixed, the individual values of $\Delta(X;3)$ and $\Delta(X;3.12)$ are not. $\Delta(X;3.12)$ quantifies the redundancy of $3$ with $12$ which is unique with respect to the redundancy of $3$ with $1$ and $2$ separately. Accordingly, to fix the value of $\Delta(X;3)$ and $\Delta(X;3.12)$ separately, we would need to also preserve the distribution of $X$ with $12$.

Importantly, $\Delta(X;3.12)$ appears alone in Eq.\,\ref{e15_2}, and not summed to $\Delta(X;3)$. Furthermore, it is the only term involving synergy of Eq.\,\ref{e15_2} which is lower in the partial ordering to node $3$ (see Figure \ref{fig0}B). This means that, to fulfill the constraint of Eq.\,\ref{e15_2}, $\Delta(X;3.12)$ can increase as much as possible  (and hence $\Delta(X;3)$ decreases, keeping their sum constant) in order to balance $\Delta(X;1.2) +\Delta(X;1.2.3)$. This is the way to minimize $I(X;123)$, since modifying $\Delta(X;3.12)$ while keeping $\Delta(X;3)+\Delta(X;3.12)$ constant has no effect on the attainable minimum within $\Delta_{1.2.3}(p)$. This freedom for adjusting $\Delta(X;3.12)$ causes that not all the redundancy $\Delta(X;1.2) +\Delta(X;1.2.3)$ will be enforced by the constraint to emerge to the synergistic terms $ \Delta(X;12) + \Delta(X;12.13) + \Delta(X;12.23) $ $ + \Delta(X;12.13.23)$. Only any part of this redundancy that cannot be balanced by $\Delta(X;3.12)$ will emerge. Note that this is only a problem when considering the constraint $C(X;1;2)=0$ within the family $\Delta_{1.2.3}(p)$, and not within $\Delta_{1.2}(p)$. The reason is that for $\Delta_{1.2}(p)$ the minimized $I(X;12)$ does not include $\Delta(X;3.12)$, and thus, if the constraint $C(X;1;2)=0$ mirrors the redundancy to $\Delta(X;3.12)$, this already makes it emerge. Conversely, for $\Delta_{1.2.3}(p)$, the term $\Delta(X;3.12)$ is part of the minimized $I(X;123)$ and thus mirroring redundancy to $\Delta(X;3.12)$ does not make it emerge.

Therefore, to make sure that the constraint $C(X;1;2)=0$ enforces the mirroring of redundancy to synergistic terms within the family $\Delta_{1.2.3}(p)$ we need an extra constraint that cancels $\Delta(X;3.12)$. This cancelation can be enforced by an extra constraint minimizing $I(X;3.12)$. The bivariate redundancy $I(X;3.12)$ is defined analogously to Eq.\,\ref{e12}d, and is equal to $\Delta(X;1.3) + \Delta(X;2.3)+$ $\Delta(X;1.2.3) + \Delta(X;3.12)$ (Figure \ref{fig0}B). From these terms, all except $\Delta(X;3.12)$ are constant within $\Delta_{1.2.3}(p)$, and thus the minimum of $I(X;3.12)$ is attained when $\Delta(X;3.12)$ vanishes. In particular, the minimum of $I(X;3.12)$ is calculated as:
\small
\begin{equation}
\min \limits_{1.2.3} I(X;3.12) = \min \limits_{1.2.3, C(X;3;12)=0} I(X;123)-\min \limits_{1.2.3} I(X;123),
\label{e15_2b}
\end{equation}
\normalsize
where $C(X;3;12)=0$ enforces the emergence of $I(X;3.12)$ but the joint distribution of $12$ and $X$ is not preserved.

Accordingly, the family of distributions in which to minimize the information $I(X;123)$ to mirror $I(X;1.2)$ to synergistic terms is
\begin{equation}
\begin{split}
&\Delta_{1.2.3, C(X;1;2),\min I(X;3.12)}(p) =  \{ q : q(x,1)= p(x,1), \\ &q(x,2)= p(x,2), q(x,3)= p(x,3),\ C(X;1;2) =0 , \\&I(X;3.12) = \min \limits_{1.2.3} I(X;3.12) \}.
\end{split}
\label{e15_3}
\end{equation}
In general, depending on the co-information constraints used, we would need to cancel the unique redundancy of a variable with the two others not only for node $3.12$ but also for $1.23$ and $2.13$, and thus we will impose also analogous constraints to these terms. For simplicity, we will refer to the set of these constraints as $s(123)$. We can then reexpress the bivariate redundancy of Eq.\,\ref{e12}d as
\small
\begin{subequations}
\begin{align}
&I(X;1.2) = \min \limits_{1.2, C(X;1;2)=0} I(X;12)-\min \limits_{1.2} I(X;12) \\
&= \min \limits_{1.2.3, C(X;1;2)=0, s(123)} I(X;123)-\min \limits_{1.2.3} I(X;123).
\end{align}
\label{e16}
\end{subequations}
\normalsize

We now consider how to separate different redundancy terms, such as the terms $\Delta(X;1.2)$ and $\Delta(X;1.2.3)$ composing $I(X;1.2)$. For this purpose, we will use conditional co-information constraints, like $C(X;1;2|3)=0$. The rationale is that conditioning removes the components of redundancy shared with the conditioning variable (dashed circles in Figure \ref{fig2}C). Furthermore, while usually conditioning also creates new synergistic contributions, if assumptions $a.1$ and $a.2$ are fulfilled, the minimization preserving only the marginals will get rid of these new contributions except to the degree in which they are required to fulfill the co-information constraints. Accordingly, in the case of imposing $C(X;1;2|3)=0$, since conditioning on $3$ eliminates $\Delta(X;1.2.3)$, only $\Delta(X; 1.2)$ is mirrored (Figure \ref{fig2}C).

Combining unconditional and conditional co-information constraints we can isolate each redundancy contribution. The measures of the trivariate decomposition are displayed in Table $1$. The use of hierarchically related constraints ensures their nonnegativity. Like for the bivariate case, it can be checked that the sum of all measures is the total mutual information $I(X;123)$. To make more apparent the hierarchical structure of the constraints, we also impose $C(X;1;2|3)=0$ when imposing $C(X;1;2)=0$. As we will explain in detail when introducing the general multivariate case, $C(X;1;2|3)=0$ has no effect once imposing $C(X;1;2)=0$. This can be intuitively understood by comparing Figure \ref{fig2}B and Figure \ref{fig2}C: in Figure \ref{fig2}B we enforce to mirror two components, while in Figure \ref{fig2}C we enforce to mirror only one of them. But enforcing to emerge something that has already been enforced to emerge has no new effect. Given $a.2$, also $s(ijk)$ has no extra effect if, like in $I(X;i.j \backslash k)$, it is not combined with an unconditional co-information, since then no constrain prevents the cancelation of the terms $i.jk$. Furthermore, although not apparent from its expression in Table $1$, $I(X;i.j.k)$ is symmetric in all variables if the assumptions $a.1$ and $a.2$ are fulfilled and, hence, the effect of the constraints can be analyzed based only on the redundancy lattice structure. Symmetry holds because the constraints could have been formulated for any permutation of the three variables, as long as the permutation was consistent between the two minimized mutual informations that are compared. Indeed, $I(X;i.j.k)$ can be reexpressed by implementing explicitly the co-informations constraints and this leads to a symmetric expression \citep{Chicharro17}.

\begin{table}
\centering
\label{t1}
\begin{tabular}{| l c | c |}
\hline \hline
Term &  Measure\\
\hline \hline
$I(X;ijk \backslash i,j,k)$ & $ I(X;ijk) - \min \limits_{i.j.k} I(X;ijk) $ \\
\hline
$I(X;i \backslash j,k)$ & $ \min \limits_{i.j.k} I(X;ijk) - \min \limits_{i.j.k} I(X;jk) $ \\
\hline
$I(X;i.j\backslash k)$ & $ \min \limits_{\begin{array}{c} {\scriptstyle i.j.k, s(ijk)}\\ {\scriptstyle C(X;i;j|k)=0} \end{array}} I(X;ijk) - \min \limits_{i.j.k} I(X;ijk) $ \\
\hline
$I(X;i.j.k)$ & $\begin{array}{c} \mathrm{min} I(X;ijk) \\ {\scriptstyle i.j.k, \ C(X;i;j)=0} \\ {\scriptstyle C(X;i;j|k)=0, s(ijk)}  \end{array} \begin{array}{c} -\\ \ \\ \  \end{array} \begin{array}{c} \mathrm{min} I(X;ijk) \\ {\scriptstyle i.j.k, s(ijk)}  \\ {\scriptstyle C(X;i;j|k)=0} \end{array}$ \\
\hline
\end{tabular}
\caption{Measures in the trivariate rooted tree decomposition. The minimization constraint $i.j.k$ preserves the marginals of each variable with the target. $C(X;i;j)=0$ and $C(X;i;j|k)=0$ constrain the co-information and conditional co-information, respectively. $s(ijk)$ indicates the set of constraints that cancel the unique redundancy of a variable with the two others (Eq.\,\ref{e15_3}).}
\end{table}

We now check how the trivariate measures implement the unfolding of the terms in the bivariate tree. We can check the unfolding $I(X;1.2) = I(X;1.2.3)+ I(X; 1.2 \backslash 3)$ combining the measures $I(X;i.j.k)$ and $I(X;i.j\backslash k)$ of Table $1$ with Eq.\,\ref{e16}b. Only we need to take into account that the constraint on the conditional co-information is subsumed by the constraint on the unconditional one, as argued above. To check the unfolding $I(X;1 \backslash  2)= I(X;1 \backslash  2,3)+I(X;1.3 \backslash 2)$ we use that
\begin{equation}
\begin{split}
I(X;1 \backslash2) &= \min \limits_{1.2} I(X;12)-\min \limits_{1.2} I(X;2) \\
&= \min \limits_{1.2.3} I(X;12)-\min \limits_{1.2.3} I(X;2).
\end{split}
\label{e17}
\end{equation}
This is because $3$ is not part of the variables for which the mutual information is being quantified. Furthermore, for $I(X;1.3 \backslash 2)$, we can explicitly enforce to the mutual information the constraint on the conditional co-information using Eq.\,\ref{e5}, that is:
\small
\begin{equation}
\begin{split}
\begin{array}{c} \mathrm{min} I(X;123) \\ {\scriptstyle 1.2.3, C(X;1;3|2)=0} \end{array} & \begin{array}{c} = \\ \  \end{array} \begin{array}{c} I(X;2) \\ \ \end{array} \begin{array}{c} + \\ \ \end{array} \begin{array}{c} \mathrm{min} I(X;13|2)\\ {\scriptstyle 1.2.3, C(X;1;3|2)=0} \end{array} \\
&= I(X;2)+ \min \limits_{1.2.3}I(X;1|2)+ \min \limits_{1.2.3}I(X;3|2),
\label{e18}
\end{split}
\end{equation}
\normalsize
where the last equality is given by the definition of the conditional co-information $C(X;1;3|2)$ as $I(X;1|2) + I(X;3|2)-I(X;13|2)$. Applying Eqs.\,\ref{e17} and \ref{e18} to compare the bivariate and trivariate measures we can check straightforwardly that the unfolding is fulfilled.

\begin{figure}
  \begin{center}
    \scalebox{0.37}{\includegraphics*{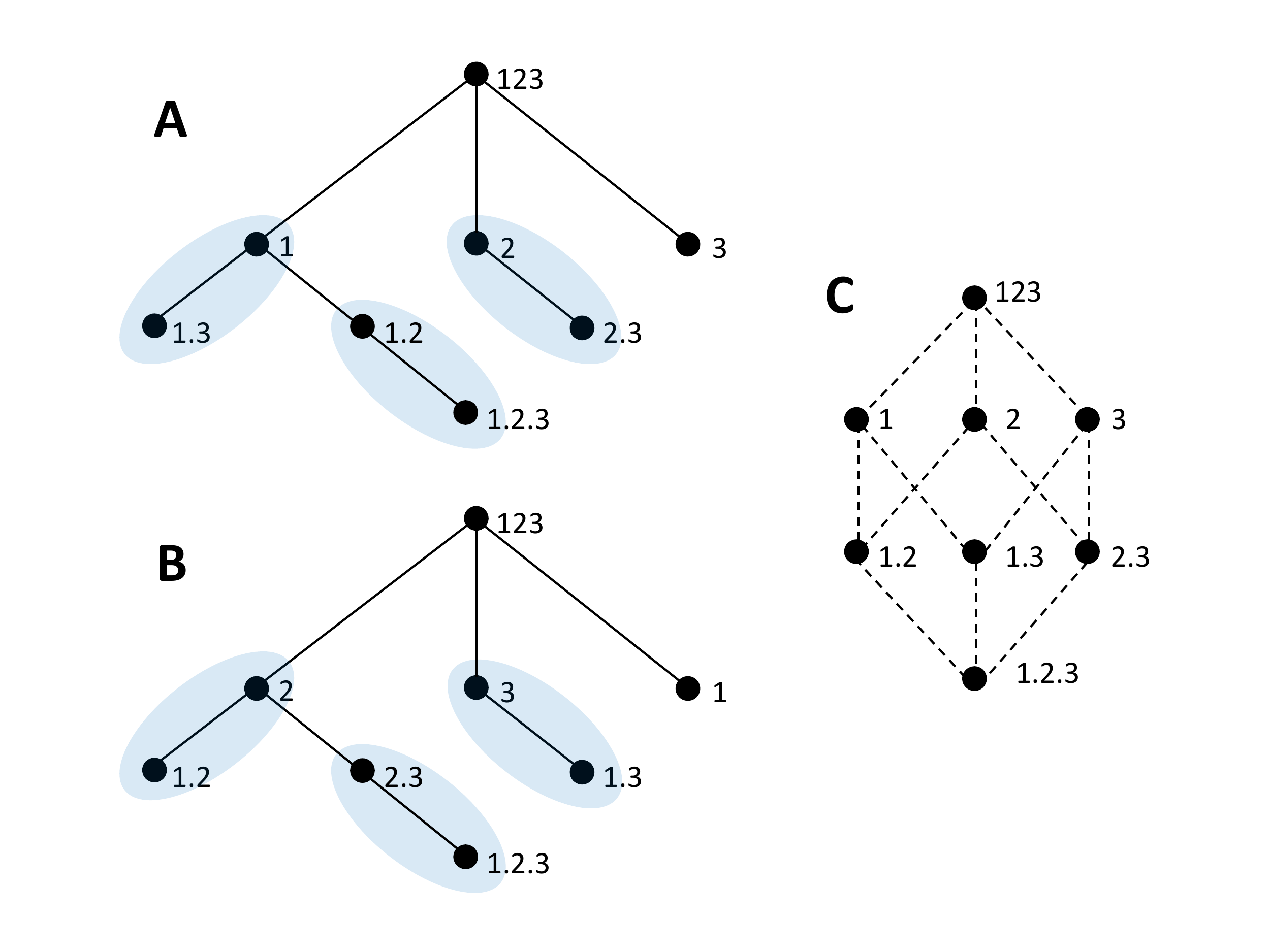}}
  \end{center}
  \caption{The association of a set of rooted trees with an adjunct lattice. \textbf{A-B)} Two alternative trees constructed from alternative orders of addition of the variables, $1 \rightarrow 2 \rightarrow 3$ and $2 \rightarrow 3 \rightarrow 1$, respectively. \textbf{C)} Adjunct lattice associated with the set of trees defined by the permutation of additions of three variables. Each edge of the lattice indicates that that edge is at least present in one of the trees.}
  \label{fig3}
\end{figure}

Like in the bivariate case, different trees can be built by permuting the order in which the variables are added. Figure \ref{fig3}A and B compare two of these trees. Although the unfolding is different, under $a.1$ and $a.2$ the measures derived remain the same because of the symmetries present in the constraints and in the mutual informations minimized. Accordingly, we can associate with the set of rooted trees defined by the permutations of the set of variables an adjunct redundancy lattice (Figure \ref{fig3}C) which reflects the partial ordering of redundancies as considered by \cite{Williams10}. The adjunct lattice has an edge if that edge is present in at least one of the trees, which means that a child term can be unfolded from its parent. Indeed, this adjunct lattice corresponds to one sublattice studied in \cite{Chicharro17}, for which generalizations of the bivariate measures of \cite{Bertschinger12} were derived following another approach, based on the connection between lattices of different order and inverting Eq.\,\ref{e8} as proposed by \cite{Williams10}. It can be checked that in fact the measures shown in Table $1$ of \cite{Chicharro17} are equal to the ones shown in Table $1$ here, in particular the trivariate redundancy is there expressed with its explicit symmetric form. However, as opposed to the complete trivariate lattice of \cite{Williams10}, with the rooted trees any synergistic component is accumulated in the root of the tree. Only the collections formed by sources containing single variables are unfolded. This implements a decomposition of the redundancy between the primary sources, but not a decomposition to separately quantify redundancy with or between synergistic contributions (e.\,g.\,$i.jk$ or $ij.ik$).

\subsection{The multivariate decomposition of redundancy in rooted trees}
\label{s4_2}

We now address in general how rooted trees decompose redundancy via local binary unfoldings with hierarchically related maximum entropy constraints. Consider a target $X$ consisting in a set of variables and a set $S$ of $n$ primary sources. For simplicity, we will from now on refer to each primary source in $S$ as a single (potentially multivariate) variable. Given $S$, we can construct $n!$ rooted trees by permuting the order in which variables are added. The set of trees associated with $S$ is defined by $\mathcal{T}(S) = \{ t \in G(n)\}$, where $G(n)$ is the group of permutations of the variables indexes, and each $t$ represents a sequence of additions to the tree. The set $\mathcal{T}(S)$ has an adjunct lattice $\mathcal{C}$ defined such that there is an edge between two nodes if and only if there is an edge between those nodes in at least one tree. We now consider how to calculate the redundancy terms associated with the adjunct lattice. For this purpose, we need to understand in general which is the effect of co-information constraints. As above, $a.1$ and $a.2$ are assumed so that the constraints can be analyzed according to the structure of the redundancy lattice. In Section \ref{s5} we will examine how the effect of the constraints changes when these assumptions do not hold.

As a first step, we generalize to the multivariate case the constraints $s(123)$ introduced when defining the family of distributions of Eq.\,\ref{e15_3}. For the multivariate case we impose that, given a set of variables $Z$, none of them has a unique redundancy with the combination of the others. To enforce this, we minimize $I(X;i.Z - i)$ $\forall i \in Z$, that is, we impose
\small
\begin{equation}
\begin{split}
\begin{array}{c} I(X;i.Z - i) \\ \\ \end{array}  \begin{array}{c} = \\ \\ \end{array} \begin{array}{c} \mathrm{min} I(X;Z) \\ {\scriptstyle m(Z) ,\ C(X;i;Z - i)=0} \end{array}  \begin{array}{c} - \\ \\ \end{array} \begin{array}{c} \mathrm{min} I(X;Z) \\ {\scriptstyle m(Z)} \end{array} \begin{array}{c} \forall i \in Z \\ \\ \end{array},
\label{e20_0}
\end{split}
\end{equation}
\normalsize
where the right-hand side expression is analogous to Eq.\,\ref{e15_2b}. In general, we will use $m(B)$ to indicate the constraint of preserving the bivariate marginals of the target with each variable in the set $B$. As before, we will use $s(B)$ to indicate the set of constraints of Eq.\,\ref{e20_0} for the variables in $B$. These two types of constraints jointly accompany the co-information constraints so that the latter produce the mirroring of redundancy to synergistic components. Accordingly, we will merge them into $ms(B)$ unless we specifically want to refer to one.

\subsubsection{The effect of co-information constraints mirroring redundancy components}
\label{s4_2_1}

If from a set of variables $Z$ we take two variables $i$ and $j$ and examine which part of the redundancy between these two variables is unique with respect to the variables in $Z' $, an exclusive subset of $Z$ with respect to $i,j$, we have that
\begin{equation}
\begin{split}
\begin{array}{c} \mathrm{min} I(X;Z) \\ {\scriptstyle ms(Z),\  c(Z')} \end{array}  &\begin{array}{c} - \\ \\ \end{array} \begin{array}{c} \mathrm{min} I(X;Z) \\ {\scriptstyle m(Z)} \end{array} \\&\begin{array}{c} = \sum \limits_{{\scriptstyle \beta \in \downarrow \alpha_{i,j;Z} - \bigcup_{k \in Z'} \downarrow \alpha_{i,j,k;Z}}} \Delta_{\mathcal{C}}(X;\beta)\\ \end{array},
\label{e20}
\end{split}
\end{equation}
where $c(B)$ stands for the constraint $C(X;i;j|B)=0$ for any set of variables $B$, and $\Delta_{\mathcal{C}}(X;\beta)$ are terms in the adjunct lattice. Here $\downarrow \alpha_{i,j;Z}$ indicates all the nodes that can be reached by descending, in the adjunct lattice formed by the $Z$ variables, from the node associated with the collection $\alpha_{i,j;Z} = i.j$. Note that only collections containing sources formed by single variables, the primary sources, appear in the lattice, other than the top collection corresponding to the root node. Furthermore, $\bigcup_{k \in Z'} \downarrow \alpha_{i,j,k;Z} $ indicates the union of the nodes that can be reached descending from at least one of the nodes $i.j.k$, for $k \in Z'$. The effect of $C(X;i,j|Z')=0$ is to mirror to the synergy all redundancy terms unique of $i$ and $j$ with respect to all the variables included in $Z'$. As argued for the trivariate case, this is because the redundancy mirrored is the redundancy of $i$ and $j$ conditioned on $Z'$. Under assumptions $a.1$ and $a.2$, the minimization constrained to preserve only the marginals (given $m(Z)$) gets rid of the synergistic components that would be created otherwise by conditioning.

We now consider how in general a combination of co-information constraints determines which redundancy components are mirrored to the synergy. Consider the set of variables $Z$, two variables $i$ and $j$ within $Z$, and $\mathbb{C} = \{Z_1,...,Z_m\}$ a set of subsets containing variables exclusive to $i,j$ in $Z$. If all sets in $\mathbb{C}$ are used to impose constraints of the form $c(Z_r)$, the redundancy contributions mirrored correspond to
\small
\begin{equation}
\begin{split}
&\begin{array}{c} \mathrm{min} I(X;Z) \\ {\scriptstyle ms(Z)\ c(Z_r)\  \forall Z_r \in \mathbb{C}}  \end{array} \begin{array}{c} - \\ \\ \end{array} \begin{array}{c} \mathrm{min} I(X;Z) \\ {\scriptstyle m(Z)} \end{array} \\ &\ \ \ \ \ \ \ \ \ \ \ \ \begin{array}{c} =  \sum \limits_{\beta \in \bigcup_{Z_r \in \mathbb{C}} \left ( \downarrow \alpha_{i,j;Z} - \bigcup_{k \in Z_r} \downarrow \alpha_{i,j,k;Z} \right)} \Delta_{\mathcal{C}}(X;\beta)\\ \\ \end{array},
\label{e21}
\end{split}
\end{equation}
\normalsize
that is, a redundancy term is mirrored if there is at least one constraint $c(Z_r)$ which enforces it to emerge. As a corollary of Eq.\,\ref{e21}, adding a second constraint with a superset $Z''$ of $Z'$, leaves invariant the minimized mutual information:
\begin{equation}
\begin{split}
\begin{array}{c} \mathrm{min} I(X;Z) \\ {\scriptstyle ms(Z) \  c(Z')} \end{array}  \begin{array}{c} =  \\ \  \end{array} \begin{array}{c} \mathrm{min} I(X;Z) \\ {\scriptstyle ms(Z),\  c(Z'), \  c(Z'')} \end{array} \begin{array}{c} \mathrm{if}\ Z' \subseteq Z'' \\ \ \end{array}.
\label{e22}
\end{split}
\end{equation}
This is because, under assumptions $a.1$ and $a.2$, as indicated by the terms comprised in the sum of Eq.\,\ref{e21}, further conditioning on a superset enforces the emergence of a subset of the already emerged redundancy terms. Eq.\,\ref{e22} indicates a hierarchical structure of the constraints, which as we will see plays an important role in the decomposition implementation.

Another special case that will be useful is that, for two exclusive subsets $Z'$ and $Z''$ of $Z$, also nonoverlapping with the variables $i$ and $j$ which redundancy is examined,
\small
\begin{equation}
\begin{split}
&\begin{array}{c} \mathrm{min} I(X;Z) \\ {\scriptstyle ms(Z)\  c(Z'',k) \forall k \in Z'} \end{array}  \begin{array}{c} - \\ \ \\ \end{array} \begin{array}{c} \mathrm{min} I(X;Z) \\ {\scriptstyle m(Z)} \end{array}  \\ & \begin{array}{c} = \sum \limits_{\beta \in \downarrow \alpha_{i,j;Z} - ((\bigcup_{k \in Z''} \downarrow \alpha_{i,j,k;Z}) \cup (\downarrow \bigwedge_{k \in Z'} \alpha_{i,j,k;Z}))} \Delta_{\mathcal{C}}(X;\beta)\\ \\ \end{array},
\label{e22b}
\end{split}
\end{equation}
\normalsize
where $\downarrow \bigwedge_{k \in Z'} \alpha_{i,j,k;Z}$ indicates the set of descending nodes from the infimum of all nodes $\alpha_{i,j,k;Z}$, $k\in Z'$, that is, it indicates all nodes that can be reached descending from the first node of intersection of all the descending paths from $\alpha_{i,j,k;Z}$, $k\in Z'$. This means that the above set of constraints leads to the emergence of all the redundancy of $i$ and $j$ also redundant to some of the  variables in $Z'$ and unique with respect to $Z''$, except the terms corresponding to the joint redundancy of all the variables in $Z'$ with $i$ and $j$ and unique with respect to $Z''$.

Using this type of constraints, we can now define multivariate measures of the redundancy of a set of variables and also of the unique redundancy of a set of variables with respect to another set.

\subsubsection{Multivariate redundancy and unique redundancy measures}

Suppose a target variable $X$ and a set of variables $S$. To explicitly consider that some of the variables of a system may not be accessible, we distinguish a subset $S' \subseteq S$ comprising only the variables observed or of interest. From $S'$ we consider a subset $Y \subseteq S'$ which unique redundancy with respect to another exclusive subset $W \subseteq S'$ is to be determined. We define $Z$ as the union $Z = Y \cup W$ and we select any pair of variables $i,j \in Y$ as reference. The unique redundancy between the variables in $Y$ with respect to the variables in $W$ is defined as:
\begin{equation}
\begin{split}
I(X;\alpha_{Y;Z}) &= \sum_{\beta \in \downarrow \alpha_{Y;S'} - \bigcup_{k \in W} \downarrow \alpha_{Y,k;S'}}\Delta_\mathcal{C}(X;\beta) \\&\begin{array}{c} = \\ \ \\ \ \end{array}  \begin{array}{c} \mathrm{min} I(X;Z) \\ {\scriptstyle ms(Z),\  c(W)}\\ {\scriptstyle c(W,k) \ \ \forall k \in Y - ij}  \end{array}  \begin{array}{c} - \\ \ \\ \ \end{array} \begin{array}{c} \mathrm{min} I(X;Z) \\ {\scriptstyle ms(Z)} \\ {\scriptstyle c(W,k) \ \ \forall k \in Y - ij}   \end{array},
\label{e23}
\end{split}
\end{equation}
where $\mathcal{C}$ is the adjunct lattice for $S'$ and $\alpha_{Y;Z}$ is the collection, in the adjunct lattice for $Z$, constituted by each single variable in $Y$ as a different source. The type of constraints $ms(\cdot)$ and $c(\cdot)$ are defined below Eqs.\,\ref{e20_0} and \ref{e20}, respectively. The sum on $\Delta (X;\beta)$ comprises those terms associated with collections that contain all variables in $Y$ but none from $W$.

We have distinguished between $S'$ and $Z$ to indicate that one can calculate a unique redundancy for a subset of $S'$ and its value will depend only on the variables in $Y$ and $W$, as if $S'=Z$. Similarly, one can take any set $Z'$ such that $Z \subseteq Z'$ and substitute $I(X;Z)$ and $ms(Z)$ by $I(X;Z')$ and $ms(Z')$ in both minimized mutual informations compared without altering the measure because only the mirrored redundancies determine its value. Furthermore, given Eq.\,\ref{e22b}, the measure is invariant for any variables $i,j \in Y$ selected as reference, and hence fulfills the symmetry axiom. Moreover, redundancy is a special case of unique redundancy, when $W$ is empty. In the special case of $Y = ij$ the constraints $c(W,k) \ \ \forall k \in Y - ij$ vanish since there is no other variable in $Y$.

Eq.\,\ref{e23} presents a way to use the co-information constraints to lead to the emergence of the desired group of redundancy contributions. However, this way is not unique. Since the group of redundancy components quantified depends only on the difference of the two minimizations, the terms mirrored for each of these minimizations can change as long as their difference is the same. Given Eq.\,\ref{e22b}, we can alternatively calculate the same redundancy as
\begin{equation}
\begin{split}
\begin{array}{c} I(X;\alpha_{Y;Z})  \\ \ \\ \ \end{array} \begin{array}{c} = \\ \ \\ \ \end{array}  \begin{array}{c} \mathrm{min} I(X;Z) \\ {\scriptstyle ms(Z),\  c(W)}\\ {\scriptstyle c(W - v,k) \ \ \forall k \in Y - ij}  \end{array}  \begin{array}{c} - \\ \ \\ \ \end{array} \begin{array}{c} \mathrm{min} I(X;Z) \\ {\scriptstyle ms(Z)} \\ {\scriptstyle c(W - v,k) \ \ \forall k \in Y - ij}   \end{array},
\label{e23b}
\end{split}
\end{equation}
where $v$ is any variable $v \in W$. We can compare these two ways in which redundancies can be estimated: In Eq.\,\ref{e23} we compare a minimization mirroring all terms descending from $i.j$ that do not have any variable from $W$ with a minimization mirroring all terms descending from $i.j$ that do not have any variable from $W$ and furthermore do not have the variables in $Y$ all together. This leads to isolate the terms descending from $i.j$ that do not have any variable from $W$ and have the variables in $Y$ all together. Given that $c(W) = c(W - v,v)$, from Eq.\,\ref{e22b} we see that in Eq.\,\ref{e23b} we compare a minimization mirroring all terms descending from $i.j$ that do not have any variable from $W - v$ and furthermore do not have ${Y,v}$ together, with a minimization mirroring all terms descending from $i.j$ that do not have any variable from $W - v$ and furthermore do not have $Y$ together. Since the same terms containing $v$ emerge in both minimizations, the measure is invariant to which $v \in W$ is selected. This comparison again leads to isolate the terms descending from $i.j$ that do not have any variable from $W$ and have the variables in $Y$ all together.

Finally, while Eqs.\,\ref{e23} and \ref{e23b} define all the redundancies and unique redundancies between at least two variables, they are not applicable to the self-redundancy and the unique information of a single variable with respect to a set of variables. This is clear since the co-information constraints used require that we can at least select two variables $i,j$ from $Y$. The self-redundancy of variable $i$ is taken to be its mutual information $I(X;i)$, which by construction then fulfills the self-redundancy axiom. The unique information for $Y=i$ with respect to $W$ is defined as:

\begin{equation}
I(X; \alpha_{i;Z}) = \min \limits_{m(Z)}I(X;Z) - \min \limits_{m(Z)}I(X;W),
\label{e24}
\end{equation}
directly generalizing the corresponding bivariate and trivariate expressions.

\subsubsection{Hierarchical decompositions in local unfoldings of redundancy}

So far we have pointed out that the redundancy measures fulfill the self-redundancy axiom by construction and the symmetry axiom because under the assumptions $a.1$ and $a.2$ the effect of the constraints of Eqs.\,\ref{e20}-\ref{e22b} is determined by the structure of the redundancy lattice. To show that the monotonicity axiom is also complied we will now examine the local binary unfoldings of information in the rooted tree. The identity axiom will be discussed in Appendix \ref{ap2}.

The measures defined above are all by definition nonnegative. In Eqs.\,\ref{e23} and \ref{e23b} nonnegativity is guaranteed because the left one of the two minimizations compared is subjected to an extra constraint, and in Eq.\,\ref{e24} the full information of $Z$ is compared with the one of $W \subseteq Z$. Nonetheless, this does not guarantee that the measures implement a decomposition of the total mutual information. To show that this is the case, we need to see that in each update of the tree, when incorporating a new variable, while the new root and new node pending from it update the synergy and add the new unique information, respectively, for all the previously existing nodes that are unfolded the unfolding preserves the information of the original term. Indeed, since each new variable contributes when added with the unique information given by Eq.\,\ref{e24}, as long as this information is preserved in subsequent unfoldings, the overall sum of redundancy terms remains the information in the marginal distributions of the whole set of variables, while all synergistic components are accumulated in the root node, and thus their sum is equal to the total mutual information.

We now show that information is preserved because at each unfolding the constraints are hierarchically related. Consider again two exclusive subsets $Y$ and $W$ of the set of observed variables $S'$ and $Z = Y \cup W$. Consider that we have calculated the term corresponding to the unique redundancy between the variables of $Y$ with respect to the variables of $W$. We now examine, when adding a new variable $v$ out of $S'$ as it is done when updating the tree, which part of the original unique redundancy is also redundant with $v$ and which part is unique. That is, we separate the unique redundancy of $\{Y,v \}$ with respect to $W$, and the unique redundancy of $Y$ with respect to $\{ W,v\}$. We see that
\begin{equation}
\begin{split}
\begin{array}{c} I(X;\alpha_{Y;Z}) \\ \ \\ \ \end{array} &\begin{array}{c} = \\ \ \\ \ \end{array}   \begin{array}{c} \mathrm{min} I(X;S') \\ {\scriptstyle ms(S') , \  c(W)}\\ {\scriptstyle c(W,k) \ \ \forall k \in Y - ij}  \end{array}  \begin{array}{c} - \\ \ \\ \ \end{array} \begin{array}{c} \mathrm{min} I(X;S') \\ {\scriptstyle ms(S')} \\ {\scriptstyle c(W,k) \ \ \forall k \in Y - ij}  \end{array} \\ &\begin{array}{c} = \\ \ \\ \ \end{array} \left [ \begin{array}{c} \mathrm{min} I(X;S') \\ {\scriptstyle ms(S') , \  c(W)}\\ {\scriptstyle c(W,k) \ \ \forall k \in Y - ij}  \end{array}  \begin{array}{c} - \\ \ \\ \ \end{array} \begin{array}{c} \mathrm{min} I(X;S') \\ {\scriptstyle ms(S'), \  c(W,v)}\\ {\scriptstyle c(W,k) \ \ \forall k \in Y - ij} \end{array}  \right ] \\ &\begin{array}{c} + \\ \ \\ \ \end{array} \left [ \begin{array}{c} \mathrm{min} I(X;S') \\ {\scriptstyle ms(S'),\  c(W,v)}\\ {\scriptstyle c(W,k) \ \ \forall k \in Y - ij}   \end{array} \begin{array}{c} - \\ \ \\ \  \end{array} \begin{array}{c} \mathrm{min} I(X;S') \\ {\scriptstyle ms(S')} \\ {\scriptstyle c(W,k) \ \ \forall k \in Y - ij}  \end{array}\right ] \\ &= I(X;\alpha_{\{Y,v\};\{Z,v\}})+I(X;\alpha_{Y;\{Z,v\}}).
\label{e25}
\end{split}
\end{equation}
The first equality comes from the definition of Eq.\,\ref{e23}. We then separate it into two components by adding and subtracting the same minimized information. The first term in the sum corresponds to the unique redundancy of $\{Y,v \}$ with respect to $W$. This can be seen from Eq.\,\ref{e23} considering that $\{ v, (Y - ij) \} = \{Y , v \}- ij$. By Eq.\,\ref{e22}, it does not matter that the constraint $c(W,v)$ is not present in the left minimization. The second term is the unique redundancy of $Y$ with respect to $\{ W,v\}$. This can be seen from Eq.\,\ref{e23b} considering that $W - v$ in Eq.\,\ref{e23b} corresponds to $W$ in Eq.\,\ref{e25}, that is, it is the conditioning set before the addition. Apart from showing how the unfolding occurs, Eq.\,\ref{e25} also shows that the redundancy measures fulfill the monotonicity axiom.

We will now also show how the unfolding works for the unique information terms of one variable as defined by Eq.\,\ref{e24}. In this case $Y=i$ and $W= Z - i$. We consider the unfolding when adding a new variable $v$. We proceed in the inverse way, showing that the sum of the redundant information of $i$ and $v$ unique with respect to $Z - i$ and the unique information of $i$ with respect to $\{ Z - i, v\}$, is equal to the unique information of $i$ with respect to $Z - i$:

\begin{equation}
\begin{split}
&I(X;\alpha_{\{i,v\};\{Z,v\}})+I(X;\alpha_{i;\{Z,v\}}) \\
&\begin{array}{c} = \\ \ \\ \ \end{array} \left [ \begin{array}{c} \mathrm{min} I(X;\{Z,v\}) \\ {\scriptstyle ms(\{Z,v\})} \\ {\scriptstyle C(X;i;v|Z - i)=0} \end{array}  \begin{array}{c} - \\ \ \\ \ \end{array} \begin{array}{c} \mathrm{min} I(X;\{Z,v\}) \\ {\scriptstyle m(\{Z,v\})}\\ \\ \end{array}  \right ] \\ &\begin{array}{c} + \\ \  \end{array} \left [ \begin{array}{c} \mathrm{min} I(X;\{Z,v\}) \\ {\scriptstyle m(\{Z,v\})} \end{array} \begin{array}{c} - \\ \ \end{array} \begin{array}{c} \mathrm{min} I(X;\{Z,v\} - i) \\ {\scriptstyle m(\{Z,v\})} \end{array}\right ] \\ &\begin{array}{c} = \\ \  \end{array}  \left [ \begin{array}{c} \mathrm{min} I(X;Z - i) \\ {\scriptstyle m(\{Z,v\})}\end{array}  \begin{array}{c} + \\ \ \\ \end{array} \begin{array}{c} \mathrm{min} I(X;v|Z - i) \\ {\scriptstyle m(\{Z,v\})} \end{array} \right. \\  &\begin{array}{c} + \\ \ \\ \end{array} \left. \begin{array}{c} \mathrm{min} I(X;i|Z - i) \\ {\scriptstyle m(\{Z,v\})} \end{array} \right ] \begin{array}{c} - \\ \ \\ \end{array} \begin{array}{c} \mathrm{min} I(X;\{Z,v\} - i) \\ {\scriptstyle m(\{Z,v\})} \end{array} \\ &\begin{array}{c} = \\ \ \end{array}  \begin{array}{c} \mathrm{min} I(X;i|Z - i) \\ {\scriptstyle m(Z)} \end{array} \begin{array}{c} = \\ \ \end{array} \begin{array}{c} I(X;\alpha_{i;Z}) \\ \ \end{array}.
\label{e26}
\end{split}
\end{equation}
The form of $I(X;\alpha_{\{i,v\};\{Z,v\}})$ comes from Eq.\,\ref{e23} and the one of $I(X;\alpha_{i;\{Z,v\}})$ from Eq.\,\ref{e24}. The term minimizing $I(X;\{Z,v\})$ under constraint $ms(\{Z,v\})$ is canceled because it appears with opposite sign in each measure. We then explicitly implement the co-information constraint and obtain $I(X;\alpha_{i;Z})$, which conforms to Eq.\,\ref{e24}.

The iterative application of these unfoldings guarantees a nonnegative decomposition of redundancy. Therefore, together with the root of the tree that captures all the synergistic contributions, a decomposition of the total mutual information is implemented. Note that since all synergistic components are accumulated in the root term, this decomposition is orthogonal to decompositions based on the hierarchical separation of higher-order moments \citep{Olbrich15, Perrone16}.

\section{Nonnegativity and the consistency of constraints}
\label{s5}

The measures of multivariate redundancy and unique redundancy have been defined based on the general effect of the minimization constraints (Eqs.\,\ref{e20_0}-\ref{e22b}) derived under assumptions $a.1$ and $a.2$. These assumptions allow analyzing the constraints only from examining the redundancy lattice. Without them, we could not establish general expressions of the effect of the constraints, and the minimizations would depend on the particular properties of each distribution. When these assumptions do not hold, the effect of the constraints generally differs from Eqs.\,\ref{e20_0}-\ref{e22b} and the preservation of information in the unfoldings is lost because Eqs.\,\ref{e25}-\ref{e26} are not fulfilled. In that case, the consistency of the measures as implementing a mutual information decomposition is lost. We now examine in more detail how the measures are affected when nonnegativity, which is assumed in $a.1$, does not hold.

To understand the effect of the existence of negative terms in the interpretability of the redundancy measures we again focus on the trivariate case. We first consider the constraints of the type using $\min I(X;3.12)$, added to define the family of distributions $\Delta_{1.2.3, C(X;1;2),\min I(X;3.12)}(p)$ of Eq.\,\ref{e15_3}. Based on Eq.\,\ref{e15_2} we argued that, if $\Delta(X;3.12)$ is allowed to be higher than zero, the minimization within the family $\Delta_{1.2.3, C(X;1;2)}(p)$ only results in the emergence to synergistic terms of the part of $\Delta(X;1.2)+ \Delta(X;1.2.3)$ that is not balanced by $\Delta(X;3.12)$. The constraint $\min I(X;3.12)$ was added to cancel $\Delta(X;3.12)$ so that all the redundancy emerges to the synergistic terms. However, this cancellation relies on the nonnegativity of $\Delta(X;3.12)$. If $I(X;3.12)$ could further be minimized in Eq.\,\ref{e15_2b} not only cancelling $\Delta(X;3.12)$ but rendering it negative then, to cancel the co-information in Eq.\,\ref{e15_2}, the synergistic terms would need to balance $\Delta(X;1.2)+ \Delta(X;1.2.3)$ and also $|\Delta(X;3.12)|$. Accordingly, the equality of Eq.\,\ref{e16}b would no longer hold. The measure in Eq.\,\ref{e16}b would still be nonnegative by construction, but would not quantify $I(X;1.2)$, in contrast to the measure in Eq.\,\ref{e16}a. Therefore, the existence of a negative term $\Delta(X;3.12)$ would be reflected in this lack of consistency.

Negative terms also affect conditional co-information constraints and their hierarchy. Consider a constraint canceling a conditional co-information, like $C(X;i;j|k)=0$ that is imposed to $I(X;ijk)$ in some of the measures of Table $1$. For example, consider the constraint
\begin{equation}
\begin{split}
&C(X;1;2|3)=0 = I(X;1|3)-I(X;1|23) \\&=\left [ \Delta(X;1.2) +\Delta(X;1.23) + \Delta(X;2.13) \right. \\&\ \ \ \ \ \  + \left. \Delta(X;13.23) + \Delta(X;12.13.23) \right]\\ &-\left [ \Delta(X;12) + \Delta(X;123) \right]
\end{split}
\label{e28}
\end{equation}
imposed to $I(X;123)$. $I(X;1|3)$ includes synergistic terms, due to the conditioning on $3$. However, as discussed in Section \ref{s4_1}, while usually conditioning creates synergistic contributions, if assumptions $a.1$ and $a.2$ are fulfilled, the minimization preserving only the marginals gets rid of them except to the degree in which they are required to fulfill the co-information constraints. That is, all terms within the first brackets except $\Delta(X;1.2)$ vanish in Eq.\,\ref{e28}, and thus to fulfill the constraint this term has to be balanced by $\Delta(X;12) + \Delta(X;123)$. However, if the synergistic terms within the first brackets could be rendered negative to balance $\Delta(X;1.2)$, the constraint would be fulfilled and $I(X;123)$ further minimized, since those terms would contribute negatively to it. Accordingly, the co-information constraint would no longer lead to the emergence of $\Delta(X;1.2)$.

Furthermore, if negative synergistic terms exist, the hierarchical relation of constraints involving subsets of conditioning variables (Eq.\,\ref{e22}) does no longer hold. To see this, we can compare the constraints $C(X;1;2|3)=0$ and $C(X;1;2|34)=0$. Given the lattice structure, the redundancy in $C(X;1;2|3)$ includes $\Delta(X;1.2)$ and $\Delta(X; 1.2.4)$, while $C(X;1;2|34)$ only includes $\Delta(X;1.2)$. As argued above, if synergistic terms can be negative, the way to minimize the mutual information (now $I(X;1234)$) when fulfilling a conditional co-information constraint is to render negative the synergistic terms added to the redundancy terms (e.\,g.\,the ones within the first brackets of Eq.\,\ref{e28}). Comparing $C(X;1;2|3)=0$ and $C(X;1;2|34)=0$, for the former negative synergistic terms need to balance $\Delta(X;1.2)+\Delta(X;1.2.4)$, while for the latter only $\Delta(X;1.2)$. That is, the negative component contributed by synergistic terms to fulfill the co-information constraint would be bigger for $C(X;1;2|3)=0$, which means that $I(X;1234)$ would be further minimized. Accordingly, in contrast to Eq.\,\ref{e22}, even if $3 \subseteq 34$ the constraint $C(X;1;2|34)=0$ would constrain more the minimization than $C(X;1;2|3)=0$.

This loss of the hierarchical relations between constraints stated in Eq.\,\ref{e22} impairs the binary unfolding of the redundancy measures (Eq.\,\ref{e25}). This is because, as we pointed out, $I(X;\alpha_{\{Y,v\};\{Z,v\}})$ is obtained by assuming that, given Eq.\,\ref{e22}, the lack of the constraint $c(W,v)$ does not alter the left minimization. Like for Eq.\,\ref{e16}b, this lack of consistency can be checked, namely by examining the match of the measures associated in each binary unfolding, or comparing the results of minimizations that add or not explicitly constraints that should not alter the minimization according to Eq.\,\ref{e22}. The effect of the constraints as described in Section \ref{s4_2_1} relies on the validity of assumptions $a.1$ and $a.2$, and hence also does the interpretation of the measures of Eqs.\,\ref{e23}-\ref{e24} as actually quantifying redundancies and unique redundancies. While these measures are nonnegatively defined by construction, their interpretability and joint consistency as implementing the decomposition is impaired when these assumptions, and in particular nonnegativity, do not hold.

\section{Discussion}

The quantification of the redundancy components in the information that a set of variables $S$ has about a target $X$ has proven to be elusive in multivariate systems. Although the mutual information decomposition into nonnegative redundant, unique, and synergistic components of \cite{Williams10} has been a fruitful conceptual framework with broad ramifications to study information in multivariate systems \citep{Lizier13, Wibral15, Banerjee15, James16, Pica17, Pica17b, Kay18}, the identification of a suited measure of multivariate redundancy is a subject of ongoing research. In particular, \cite{Harder12} indicated that the original measure proposed by \cite{Williams10} only quantifies common amounts of information. Subsequent proposals either focus on the bivariate case \citep{Harder12, Bertschinger12}, do not require nonnegativity \citep{Ince16, Finn18}, or focus on characterizing the synergistic components of information \citep{Griffith13, Perrone16}.

From all proposed measures, the maximum entropy measures have a preeminent role in the bivariate case because, under certain assumptions, they provide bounds for, or match, the actual terms of the decomposition \citep{Bertschinger12}. Motivated by this especial role, we have generalized the maximum entropy approach proposing definitions of the redundant information that a group of variables has about a target, and of the unique redundant information that a group of variables has about a target that is not redundant with information from another group. These quantities are embedded in rooted tree decompositions of the mutual information, based on the local unfolding of redundancy components when a new variable is added to the tree. We have shown that each redundancy component can be decomposed into a component also redundant with the new variable and a component of unique redundancy with respect to it. This unfolding is implemented with hierarchically related maximum entropy constraints, which guarantees the nonnegativity of all the terms in the decomposition.

In Section \ref{s3} we revisited the bivariate maximum entropy measures of redundant, unique, and synergistic information \cite{Bertschinger12}. We showed that these measures can be reexpressed as implementing binary unfoldings of the mutual information and conditional mutual information. In particular, we identified the entropy maximization constraint that allows quantifying redundancy. This constraint enforces the cancellation of the co-information \citep{Bell03} -which has been shown by \cite{Williams10} to quantify the difference between redundancy and synergy- while preserving only the marginal distributions of each variable in $S$ with the target. This has the effect of mirroring the redundancy into the otherwise minimized synergistic component.

To generalize this approach to multivariate systems, we showed that the especial role of the maximum entropy measures in connection to the actual decomposition also holds for the multivariate case, under assumptions analogous to the ones of \cite{Bertschinger12}, which comprise the assumption that a nonnegative decomposition exists (Section \ref{s4_0}). 
We then considered first the trivariate case (Section \ref{s4_1}) and showed how to isolate specific components of redundancy with constraints on conditional co-informations. In Section \ref{s4_2} we presented the general development of the multivariate redundancy decompositions. We provided general formulas of the effect of constraints on co-information and conditional co-information, under the assumptions that link the maximum entropy and actual decomposition. We then derived multivariate measures of redundancy and unique redundancy, and showed that they implement a nonnegative rooted tree decomposition of the mutual information. In Section \ref{s5} we examined how the interpretability of these measures is affected if the nonnegativity of the terms in the decomposition does not hold \citep{Bertschinger12b, Chicharro17c}. We showed that in this case the measures do not implement a mutual information decomposition because the relations between different co-information constraints change.

In our approach, the selection of the co-information constraints used to isolate different redundancy components relies on the redundancy axioms and on the partial ordering of redundancy terms introduced by \cite{Williams10}. However, the rooted tree decompositions only separate the components of redundancy and unique redundancy between the primary sources in $S$, but do not break down contributions that involve unique redundancy with sources comprising several primary sources \citep{Williams10}. Furthermore, all synergistic components are accumulated in the root term of the tree, so that these decompositions are orthogonal to decompositions hierarchically separating high-order moments \citep{Olbrich15, Perrone16}.

Regarding future extensions, a case which deserves special attention is the application of rooted tree decompositions to study dynamic dependencies in multivariate systems \citep{Chicharro12b, Faes15, Ay15}. The work in \cite{Williams11} applied the original measures of \cite{Williams10} to decompose a particular conditional mutual information, namely Transfer Entropy \citep{Marko73, Schreiber00b}, which quantifies information transfer in dynamical processes. This decomposition allows separating state-independent and state-dependent components of information transfer, and also identifying the information transferred about a specific variable \citep{Beer15}. This breakdown of Transfer Entropy could alternatively be implemented using the maximum entropy measures. More generally, it is an open question whether these methods to characterize synergy and redundancy can be combined with an interventional approach suited to quantify causal effects \citep{Ay06, Lizier10, Chicharro12}.

Multivariate measures of redundancy can be useful in many domains of data analysis, like model selection \citep{Burnham2002} or independent component analysis \citep{Hyvarinen01}. As an example of their relevance in a concrete field, we consider several applications in systems and computational neuroscience. The characterization of redundancies both in sensory stimuli and in neural responses is a fundamental step towards understanding sensory neural representations and their processing. Regarding redundancies in the stimuli, it has been a long-standing hypothesis that the brain adapts to the statistics of natural stimuli to optimize sensory processing \citep{Barlow01, Olshausen96}. While in this context redundancies may be seen in terms of the joint entropy of the multivariate stimuli, it has been argued that efficient coding must also take into account the goal of the sensory representation \citep{Rothkopf09}. Accordingly, for example for predictive coding \citep{Palmer15}, redundancies should be assessed in relation to a target variable associated with the goal. Regarding redundancies in the neural responses, a decomposition of redundancy terms can help to identify how neural representations are distributed in different neural features and different spatial and temporal scales \citep{Panzeri10, Panzeri15} as well as to understand how sensory and behavioural information is combined in neural responses \citep{Panzeri17}. Furthermore, as a particular application of the information transfer measures, the analysis of information flows between brain areas to characterize dynamic functional connectivity \citep{Valdes11, Chicharro14} can also benefit from the multivariate redundancy measures to further determine the degree of functional integration and segregation of neural dynamics \citep{Deco15}.

As discussed above, the possibility to systematically characterize redundancy in multivariate systems is expected to have applications in many domains, comprising the study of biological and social systems. This work proposed multivariate redundancy measures within the maximum entropy framework, generalizing the bivariate decomposition of \cite{Bertschinger12}. 
Given the especial link between the maximum entropy and actual decompositions, we expect these measures to be useful in practice for many multivariate systems. Future research will be required to extend to the multivariate case the efficient algorithms developed to estimate the maximum entropy decomposition in the bivariate case \citep{Makkeh17, Banerjee17, Makkeh18}. However, we have also indicated how the interpretation of the multivariate maximum entropy redundancy measures can be impaired in the presence of negative terms, and how to check the consistency of the decomposition. As pointed out in \citep{Chicharro17c}, the maximum entropy approach assumes a certain criterion to identify pieces of information, based on target-sources variables associations, which is incompatible with ensuring nonnegativity, e.\,g.\,in the presence of deterministic target-source dependencies. It remains an open question whether the lack of a nonnegative decomposition is a signature of certain systems, or if a meaningful nonnegative decomposition could be obtained with a different criterion of information identity \citep{Chicharro17c}.

\begin{acknowledgments}

This work was supported by the Fondation Bertarelli. I am grateful to S. Panzeri, G. Pica, and E. Piasini for useful discussions on these topics.

\end{acknowledgments}

\appendix

\section{Proofs of the relation between the maximum entropy and actual terms of the decomposition}
\label{ap}

We here prove Lema $l.1$. For that purpose, we start extending to the multivariate case the results on how synergy is bounded by its maximum entropy estimator that were stated for the bivariate case in Lema $3$ of \cite{Bertschinger12}.
\vspace*{2mm}

\noindent \textbf{Lemma l.2}: \emph{Consider a target $X$ and a set $S$ of $n$ primary sources with a distribution $p(X,S)$. Consider the family of distributions $\Delta_{S_1.S_2.....S_n}(p)$ preserving the marginal distributions of $X$ with each primary source. Assume that an actual decomposition of the mutual information exists for each distribution within the family, conforming to assumption $a.1$. Define the total synergy of $S$ with respect to the separate primary sources as}
\begin{equation}
I(X; S \backslash S_1,...,S_n)= \sum_{\beta \in (\bigcup \downarrow \alpha_{S_i})^C}  \Delta(X;\beta)
\label{e29}
\end{equation}
\emph{where $(\bigcup \downarrow \alpha_{S_i})^C$ indicates all nodes that cannot be reached descending from at least one of the nodes $\alpha_{S_i} = S_i$ corresponding to the primary sources. Then:}
\begin{equation}
I_0(X; S \backslash S_1,...,S_n)\geq I(X; S \backslash S_1,...,S_n), \notag
\label{e30}
\end{equation}
\emph{where $I_0$ belongs to the actual decomposition and $I$ is the maximum entropy estimator.}
\vspace*{2mm}

\noindent \textbf{Proof:} The proof is analogous to the one of Lema $3$ of \cite{Bertschinger12}. Define the maximum entropy measure $I(X; S \backslash S_1,...,S_n)$ as
\begin{equation}
I(X; S \backslash S_1,...,S_n)= I(X;S)- \min \limits_{S_1.S_2.....S_n} I(X;S).
\label{e31}
\end{equation}
By assumption $a.1$, within the family $\Delta_{S_1.S_2.....S_n}(p)$ the synergy $I_0(X; S \backslash S_1,...,S_n)$ is nonnegative and the nonsynergistic component $I(X;S)-I_0(X; S \backslash S_1,...,S_n)$ is invariant for all distributions. Because for all distributions within the family $I(X;S)$ is the sum of the nonnegative synergistic term and the invariant nonsynergistic term, considering this sum for the distribution that minimizes $I(X;S)$ leads to
\begin{equation}
I(X;S)-I_0(X; S \backslash S_1,...,S_n) \leq \min \limits_{S_1.S_2.....S_n} I(X;S),
\label{e32}
\end{equation}
and accordingly,
\begin{equation}
I_0(X; S \backslash S_1,...,S_n) \geq I(X;S)- \min \limits_{S_1.S_2.....S_n} I(X;S),
\label{e33}
\end{equation}
which given Eq.\,\ref{e31} proves the inequality.$\ \ \ \ \ \ \  \Box$
\vspace*{2mm}

We can now proceed with the proof of Lema $l.1$.
\vspace*{2mm}

\noindent \textbf{Proof of l.1}: We start proving that if there is a $q \in \Delta_{S_1.S_2.....S_n}(p)$ for which all synergistic terms vanish, then the redundancy maximum entropy measures retrieve the corresponding redundancy terms of the actual decomposition. We provide first a detailed proof for the trivariate case building on the results of \cite{Bertschinger12} for the bivariate one.

If all synergistic terms vanish in the trivariate lattice for at least one distribution $q'$, for $q'$ they also vanish in any bivariate lattice related to the trivariate one (Figure \ref{fig0}). Accordingly, based on Lema $3$ of \cite{Bertschinger12}, the maximum entropy and actual decompositions coincide for all associated bivariate decompositions. Since in the rooted tree decompositions all synergistic terms are accumulated in the root node, we focus in the trivariate sublattice corresponding to the adjunct lattice of Figure \ref{fig3}C, instead of considering the complete trivariate lattice of Figure \ref{fig0}B. In this sublattice, $\Delta(X;i)$ is equal to $\Delta(X;i)+\Delta(X;i.jk)$ of the complete trivariate lattice, while all the redundancy terms are equal. Given the mapping of terms between bivariate and trivariate lattices (Figure \ref{fig0}), for $q'$ the equality between the actual and maximum entropy decompositions for the bivariate lattices results in the following equalities for the terms of the sublattice: \small
\begin{subequations}
\begin{align}
&\Delta_0(X;i.j)+\Delta_0(X;i.j.k) = \Delta(X;i.j)+\Delta(X;i.j.k) \\
& \Delta_0(X;i)+\Delta_0(X;i.j) = \Delta(X;i)+\Delta(X;i.j).
\end{align}
\label{e34}
\end{subequations} \normalsize
For $q'$ the top term of the sublattice vanishes, since $\Delta_0(X;ijk)=I_0(X; S \backslash S_1,...,S_n)=0$. By lemma $l.2$, $\Delta_0(X;ijk)= 0 \geq \Delta(X;ijk)$. Since $\Delta(X;ijk)$ is nonnegative (Eq.\,\ref{e31}), this implies $\Delta_0(X;ijk)= \Delta(X;ijk)$ for $q'$. Now, by construction of the sublattice
\begin{equation}
\begin{split}
&I(X;ij|k) = \Delta(X;j)+\Delta(X;i)+\Delta(X;i.j)+\Delta(X;ijk)\\
&= \Delta_0(X;j)+\Delta_0(X;i)+\Delta_0(X;i.j)+\Delta_0(X;ijk).
\end{split}
\label{e35}
\end{equation}
Given Eq.\,\ref{e34}b and $\Delta_0(X;ijk)= \Delta(X;ijk)$, this means that $\Delta_0(X;j)= \Delta(X;j)$ for $q'$. Using this equality in Eq.\,\ref{e34} we establish the equality for the rest of redundancy terms for $q'$. Furthermore, since nonsynergistic terms are assumed to be invariant within the family, these equalities hold for all distributions within the family.

The proof for the multivariate case proceeds analogously to the one for the trivariate case, using induction. Given that Lemma $l.1$ holds for the case of $n-1$ primary sources, we can show that it also holds for $n$ sources. In particular, like in Eq.\,\ref{e34} (where $n-1=2$), in general the equalities for single terms of the actual and maximum entropy decompositions for $n-1$ lead to equalities between sums of two terms for $n$, given the binary unfolding in the rooted trees of a term into two terms when adding a new variable. Furthermore, the equality of Eq.\,\ref{e35} is valid generically for $I(X;ij |S - ij)$, being $S$ the set of $n$ primary sources. For the trivariate case this results in $I(X;ij|k)$ because $S - ij = k$. Therefore, the combination of general equations analogous to Eqs.\,\ref{e34} and \ref{e35} leads to the equality between the maximum entropy and actual decompositions for the general multivariate case when at least for one distribution $I_0(X; S \backslash S_1,...,S_n)=0$.

We now prove the converse part of the lemma. If equality between the maximum entropy and actual terms holds for all redundancy and unique redundancy terms then $I(X;S)-I_0(X; S \backslash S_1,...,S_n)$ is equal to $I(X;S)-I(X; S \backslash S_1,...,S_n)$, which means that also $I_0(X; S \backslash S_1,...,S_n) = I(X; S \backslash S_1,...,S_n)$. Now, by definition of $I(X; S \backslash S_1,...,S_n)$ in Eq.\,\ref{e31}, there is a distribution within the family for which it vanishes, namely the one with minimal information. $\ \ \ \ \ \Box$

\section{The fulfillment of the constraints in the presence of deterministic target-sources dependencies}
\label{ap2}

We here follow \citep{Chicharro17c} and examine how deterministic target-sources dependencies affect the multivariate redundancy measures. As a particular subcase, we examine if the reexpressed maximum entropy redundancy measure (Eq.\,\ref{e12}d) conforms to the identity axiom, which concerns bivariate systems in which the target is a copy of the two sources. More generally, the case of deterministic target-sources dependencies is interesting because of their role causing negative terms in the decomposition \citep{Bertschinger12b, Chicharro17c}.

In particular, we here consider the effect on the co-information constraints of the existence of some source variables that are also part of the target. Consider the conditional co-information $C(X;i;j|k)$. From the definition of the co-information (Eq.\,\ref{e5}), if there is a subset of variables $V$ overlapping between $X$ and $i$, i.\,e.\,$V \subseteq i$, $V \subseteq X$, then
\begin{equation}
\begin{split}
C(X;i;j|k) &= C(X - Vk; i - Vk; j - Vk | Vk) \\&+ I(V - k; j - k |k).
\end{split}
\label{e27}
\end{equation}
This equality can be applied iteratively when there are several overlaps between $X$, $i$, and $j$. If $Vk$ subsumes $X$, $i$, or $j$, then $C(X;i;j|k)$ equals $I(V - k; j - k |k)$. For systems for which $C(X;i;j|k)$ equals a mutual information, it may not be possible to fulfill the constraints of the form $C(X;i;j|k)=0$ used to calculate the multivariate redundancy measures. This occurs if $I(V - k; j - k |k)>0$ and is constant within the family where the minimization is performed.

For example, consider a target $X=X'12$ that includes a copy of the sources $1$ and $2$ and other variables contained in $X'$. The subcase $X' = \emptyset$ corresponds to the identity axiom. Given Eq.\,\ref{e27}, $C(X;1;2) = I(1;2)$. Since the family $\Delta_{1.2}(p)$ preserves $p(X,1)$ and $p(X,2)$, when $X$ includes at least one of the two sources $I(1;2)$ is constant within this family, and so it is $C(X;1;2)$. Given the original bivariate definition of redundancy of \citep{Bertschinger12} (Eq.\,\ref{e11}c), this leads to $I(X;1.2)= I(1;2)$, which conforms to the identity axiom. However, if directly using the alternative definition of bivariate redundancy (Eq.\,\ref{e12}d), the co-information constraint cannot be fulfilled when $I(1;2)$ is nonzero.

To get around this limitation, when the co-information constraints cannot be fulfilled because of target-sources deterministic dependencies, the calculation of the maximum entropy redundancy measures should be preceded by the separation of stochastic and deterministic components of the redundancy terms, as analyzed in \citep{Chicharro17c}. We have not addressed this refinement of the definitions here since, although theoretically relevant, the case of deterministic target-sources dependencies has a narrow scope in practice. Briefly, the same definitions of the multivariate redundancy measures can be applied to the stochastic components of the redundancy terms, while the deterministic components are calculated separately.

%
%

\end{document}